\documentclass{aa}  

\usepackage{graphicx}
\usepackage{txfonts}
\usepackage{subfig}             
\usepackage{epstopdf}           

\def\ltsima{$\; \buildrel < \over \sim \;$}
\def\simlt{\lower.5ex\hbox{\ltsima}}
\def\gtsima{$\; \buildrel > \over \sim \;$}
\def\simgt{\lower.5ex\hbox{\gtsima}}

\def\ergs{{erg s$^{-1}$}}
\def\cm2{{cm$^{-2}$}}

\def\lum{{$L_{\rm X}$}}

\def\p1{{Paper I}}

\def\xmm{{\em XMM--Newton}}
\def\chandra{{\em Chandra}}

\def\xmm{{\em XMM--Newton}}

\def\nh{{$N_{\rm H}$}}

\def\xray{{X--ray}}
\def\f14{{10$^{-14}$}}
\def\f13{{10$^{-13}$}}
\def\f12{{10$^{-12}$}}
\def\f11{{10$^{-11}$}}

\def\4u{{4U~1344$-$60}}

\def\lir{{$L_{\rm IR}$}}
\def\lirsf{{$L_{\rm IR}^{\rm SF}$}}

\def\lbol{{$L_{\rm Bol}$}}

\def\herschel{{\it Herschel}}
\def\msun{{$M_{\rm \odot}$}}

\def\mstar{$M_*$}
\def\mbh{$M_{BH}$}

\def\edd{$\lambda_{\rm Edd}$}

\usepackage{color}
\usepackage{amstext}

\begin{document}

\title{AGN vs. host galaxy properties in the COSMOS field}
           

   \author{G. Lanzuisi\inst{1,2}\thanks{\email{giorgio.lanzuisi2@unibo.it}} 
   \and
   I. Delvecchio\inst{3}
   \and
   S. Berta\inst{4}\thanks{Visiting scientist}
   \and
   M. Brusa\inst{1,2}
    \and
    A. Comastri\inst{2}
    \and
   R. Gilli\inst{2}
    \and
   C. Gruppioni\inst{2}
    \and
    S. Marchesi\inst{2,5}
    \and
   M. Perna\inst{1,2}
   \and
   F. Pozzi\inst{1,2}
    \and
   M. Salvato\inst{6}
    \and
   M. Symeonidis\inst{7}
   \and
   C. Vignali\inst{1,2}
   \and
   F. Vito\inst{8}
    \and
    M. Volonteri\inst{9}
    \and
   G. Zamorani\inst{2}.
   }

 \titlerunning{AGN vs. host properties in COSMOS}\authorrunning{G.~Lanzuisi et al.}


\institute{Dipartimento di Fisica e Astronomia, Universit\`a  di Bologna, Viale Berti Pichat 6/2, I-40127 Bologna, Italy \and
INAF - Osservatorio Astronomico di Bologna,  Via Ranzani 1, I--40127 Bologna, Italy \and 
Department of Physics, University of Zagreb, Bijeni\v{c}ka cesta 32, HR-10002 Zagreb, Croatia\  \and
University of Zagreb, Physics Department, Bijeni\v{c}ka cesta 32, 10002 Zagreb, Croatia \and
Department of Physics \& Astronomy, Clemson University, Clemson, SC 29634, USA\and
Max-Planck-Institut f\"ur extraterrestrische Physik,  Giessenbachstrasse, 85748 Garching, Germany \and
Mullard Space Science Laboratory, University College London, Holmbury St. Mary, Dorking, Surrey RH5 6NT, UK\and
Department of Astronomy and Astrophysics, 525 Davey Laboratory, The Pennsylvania State University, U. Park, PA 16802, USA \and 
Institut d'Astrophysique de Paris, UPMC et CNRS, UMR 7095, 98 bis bd Arago, F-75014 Paris, France
}
   \date{Received 25 October 2016; Accepted 22 February 2017}

\abstract
{  
The coeval active galactic nuclei (AGN) and galaxy evolution and the observed local relations between
super massive black holes (SMBHs) and galaxy properties suggest
some sort of connection or feedback between the SMBH growth (i.e., the AGN activity) and the galaxy build-up (i.e., 
the star formation history).}
{We have looked for correlations between average properties of X-ray detected AGN and their 
FIR detected, star forming host galaxies, in order to find quantitative evidences for this connection,
that has been highly debated in the latest years.}
{We exploit the rich multi-wavelength data set (from X-ray to FIR) available in the COSMOS field for a large sample
(692 sources) of AGN and their hosts, in the redshift range $0.1<z<4$. We use X-ray data to select AGN
and determine their properties, such as X-ray intrinsic luminosity and nuclear obscuration,
and broad-band (from UV to FIR) SED fitting results to derive host galaxy properties, such as 
stellar mass (\mstar) and star formation rate (SFR).}
{We find that the AGN 2-10 keV luminosity (\lum) and the host $8-1000~\mu m$ star formation luminosity (\lirsf) 
are significantly correlated, even after removing the dependency of both quantities with redshift.
However, the \textit{average} host \lirsf\ has a flat distribution in bins of AGN \lum, 
while the \textit{average} AGN \lum\ increases in bins of host \lirsf\ with logarithmic slope of $\sim0.7$,
in the redshifts range $0.4<z<1.2$.
We also discuss the comparison between the full distribution of these two quantities 
and the predictions from hydrodynamical simulations. 
No other significant correlations between AGN \lum\ and host properties is found.
On the other hand, we find that the \textit{average} column density (\nh) 
shows a clear positive correlation with the host \mstar, at all redshifts, 
but not with the SFR (or \lirsf). This translates into a negative correlation with specific SFR, at all redshifts. 
The same is true if the obscured fraction is computed.}
{Our results are in agreement with the idea, introduced in recent galaxy evolutionary models, 
that BH accretion and SF rates are correlated, but 
occur with different variability time scales. 
Finally, the presence of a positive correlation between \nh\ and host \mstar\ suggests that 
the column density that we observe in the X-rays is not entirely due to the circum-nuclear obscuring torus,
but may also include a significant contribution from the host galaxy.
}

   
   \maketitle
%

\section{Introduction}
\label{sec:intro}

Super massive black hole (SMBH) growth and galaxy build up follow a similar evolution through cosmic history, 
with a peak at $z\sim2-3$ and a sharp decline toward the present age (see Madau \& Dickinson 2014 for a review).
Furthermore, at $z=0$, SMBH and their hosts sit on tight relations that link the SMBH mass
and the bulge properties of the host, such as luminosity, stellar mass and velocity dispersion 
(Kormendy \& Richstone 1995, Magorrian et al. 1998, Kormendy \& Ho 2013).
Therefore the SMBH growth and the star formation history are likely related in some way during the cosmic time.

The key parameter that regulates both processes seems to be cold/molecular gas supply (Lagos et al. 2011, 
Vito et al. 2014, Delvecchio et al. 2015, Saintonge et al. 2016).
Star formation-related processes (e.g. supernova and stellar wind) are known to produce galaxy-wide outflows
that can regulate the in-fall of gas and therefore the star formations itself (e.g. Genzel et al. 2011). 
But more powerful mechanisms, globally identified as `AGN feedback', have been invoked 
in numerical and semi-analytic models of galaxy evolution (e.g. Granato et al. 2004; Di Matteo, Springel
\& Hernquist 2005, Menci et al. 2008, Sijacki et al. 2015, Dubois et al. 2016, Pontzen et al. 2016)
in order to reproduce the observed galaxy population, and particularly the high mass end of the galaxy
mass function.

Observationally, the role of AGN activity in influencing the evolution of the global galaxy population is not clear yet.
This issue has been investigated, in the past, looking for correlations between average AGN and host properties, 
such as BH accretion rates (BHAR) or AGN luminosity (typically in the $2-10$ keV band, \lum\ hereafter)
on one side, and SF rates (SFR) or IR luminosity (in the $8-1000~\mu m$, \lir\ hereafter) on the other side, 
taking advantage of the wealth of multi-wavelength information collected in
deep extragalactic surveys. However, different, somewhat contradictory results have been reported in the past years.

Several studies have found a flat distribution computing average \lir\ in bins of \lum\ of 
X-ray selected sources (or SFR and BHAR, respectively, 
e.g. Shao et al. 2010, Rovilos et al. 2012, Rosario et al. 2012)
at low redshift and luminosities. A significant positive correlation instead 
appears for luminous AGN (\lum$>10^{43-44}$ \ergs) and high redshifts ($z>1-2$),
suggesting two different triggering mechanisms
at high and low luminosities, via merger and secular evolution, respectively.

Other groups have found a linear correlation
at all z and \lum, when computing the average \lum\ in bins of \lir\  (in log-log space) of IR selected sources
(Chen et al. 2013, Delvecchio et al. 2015, Dai et al. 2015), with the 
ratio Log(SFR/BHAR)$\sim3$, roughly consistent with the local M$_{bulge}$/SMBH value (Magorrian et al. 1998, 
Marconi \& Hunt 2003).
Finally, other authors have found no correlation at all, regardless of the \lum\ and z range
(e.g. Mullaney et al. 2012, Stanley et al. 2015).

Looking at AGN obscuration, Rovilos et al. (2012) explored for the first time the possible relation between 
AGN column density (\nh), as measured from the X-ray spectra, and host properties, finding no correlation on a sample of 65 sources
in the XMM-CDFS survey (Comastri et al. 2011).
Rosario et al. (2012) found similar result from hardness ratios (HR) on a larger sample in COSMOS, while Rodighiero el al. (2015) 
found a positive correlation between \nh\ and \mstar, again based on HR, on a sample of $z\sim2$ AGN hosts in the same field.

From a technical point of view, these differences may partly arise from different biases and analysis methods.
For example, given that only a small fraction of the X-ray detected sources are FIR detected, and vice-versa, 
most of these studies rely on X-ray or FIR staking in order to recover the average properties of large samples of
AGN/host systems, or are limited to small subsamples.
Mullaney et al. (2015) pointed out that modeling the SFR distribution of X-ray selected hosts
as a log-normal distribution, and including upper-limits, gives different results than computing the linear mean of 
the distribution (i.e. via staking), that is instead driven upwards by the bright outliers.

Another issue was raised by Symeonidis et al. (2016), showing that the intrinsic 
AGN SED in the FIR is cooler than usually assumed.
Therefore in some cases there is no `safe' photometric band which can be used in calculating the SFR, 
without subtracting the AGN contribution. On the other hand, several of the works cited above, take directly the FIR photometry 
(typically at $60\mu m$) in order to estimate SFR, thus potentially introducing a spurious correlation at high AGN luminosities.

Recently, from theoretical studies, a physical mechanism has been proposed to explain part of these 
contradictory results: Volonteri et al. (2015a,b) explain these different observations with the way we analyze the data: 
the bivariate distribution of AGN and SF luminosities gives two very different results, depending on the binning axis. 
Hickox et al. (2014) reached similar conclusions starting from the simple assumptions that long term AGN 
activity and SFR are perfectly correlated, that the observed SFR is the average over $\sim100$ Myr, while the 
AGN activity, traced by X-ray emission, varies on a much shorter time scales. 
In these models the different time scales involved in AGN and SF variability wash out the linear dependency between 
the two quantities, if the rapidly variable AGN luminosity is used to build the subsamples to be studied `on average'. 
This result was also confirmed observationally by Dai et al. (2015) using shallow data from XMM-LSS.

Furthermore, Volonteri et al. (2015a) suggest that spatial scales are important:
the BH accretion rate should be correlated with the nuclear ($<100$~pc) SFR, 
while it is less correlated with the total ($<5$~kpc) SFR, 
except for the most intense merger episodes,
that are able to affect the whole host galaxy. Of course, the SFR that can be inferred from the FIR luminosity is 
the global, galaxy-scale SFR (with the exception of the local Universe, see e.g. Diamond-Stanic \& Rieke 2012), 
and this introduce another source of uncertainty in the observational comparison between BHAR and SFR.

Here we explore the possible correlations between AGN and host properties for a large sample of X-ray and FIR detected 
sources thanks to the extensive \chandra, \xmm\ and \herschel\ coverage on the COSMOS field (Scoville et al. 2007, 
Hasinger et al. 2007, Elvis et al. 2009, Lutz et al. 2011, Oliver et al. 2012).
This approach avoids the uncertainties related to the staking, and allow for a proper SED deconvolution, source by source.
This of course limits the significance of our findings to the brightest,
most accreting and most star forming systems. These are, however, the most interesting ones: the ones for 
which there is less agreement in the literature on the presence of a correlation between AGN and SF, 
and also the ones for which theoretical models predict that the correlation should be stronger.

The paper is organized as follows: section 2 describes the sample and source properties; section 3
presents the analysis of \lum\ and \lir\ distributions; in section 4 we compare our results
with a set of hydrodynamical simulations; in section 5 we discuss  correlations between nuclear obscuration
and host properties and in section 6 we discuss our results.
Throughout the paper, we assume a
standard $\Lambda$CDM cosmology with
$H_0=70$ km s$^{-1}$ Mpc$^{-1}$, $\Omega_\Lambda=0.73$ and $\Omega_M=0.27$ (Bennett et al 2013).

\section{The sample}

 \begin{figure*}[t]
 \begin{center}
 \includegraphics[width=8cm]{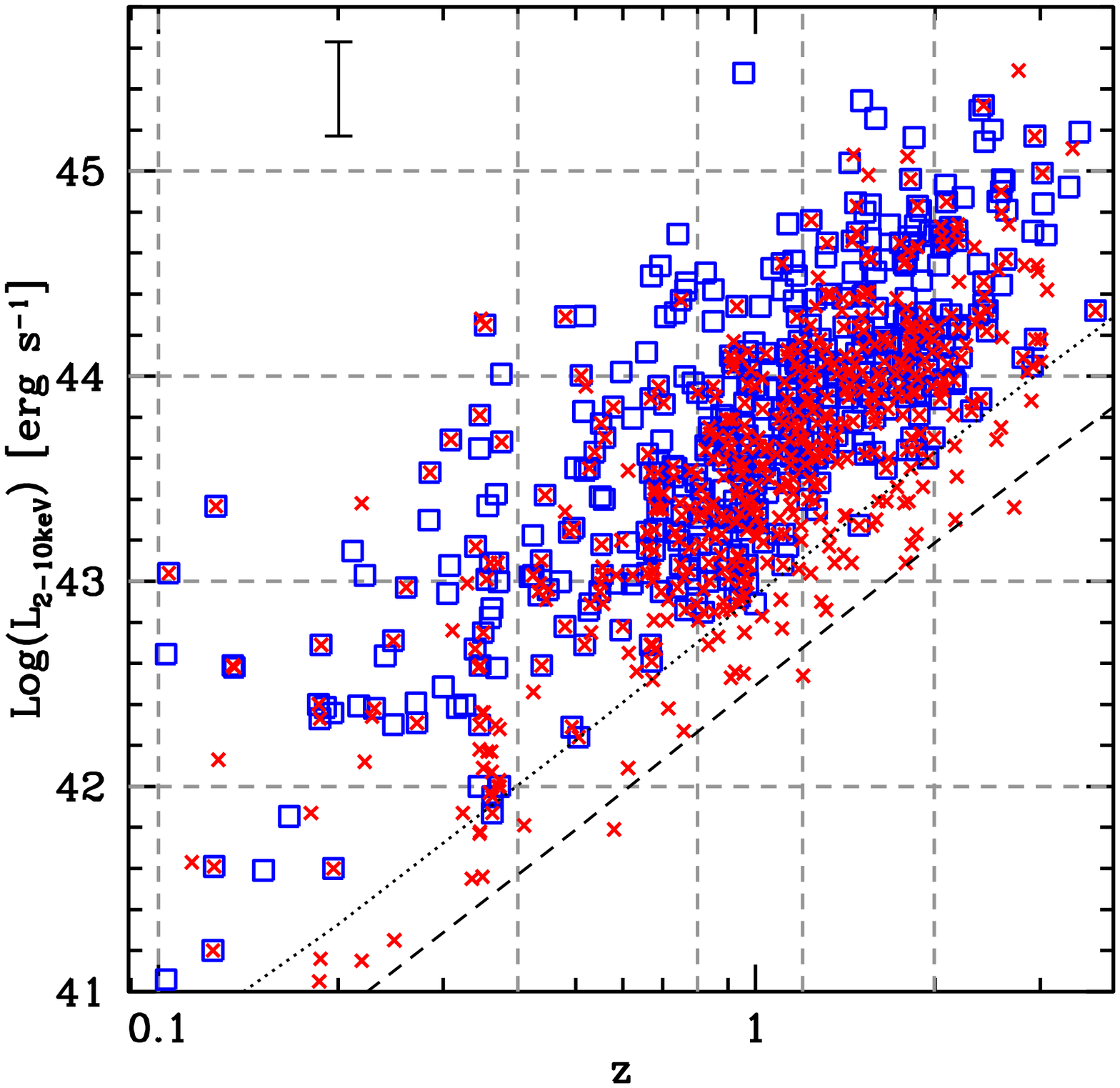}
 \hspace{1cm}
 \includegraphics[width=8cm,height=8cm]{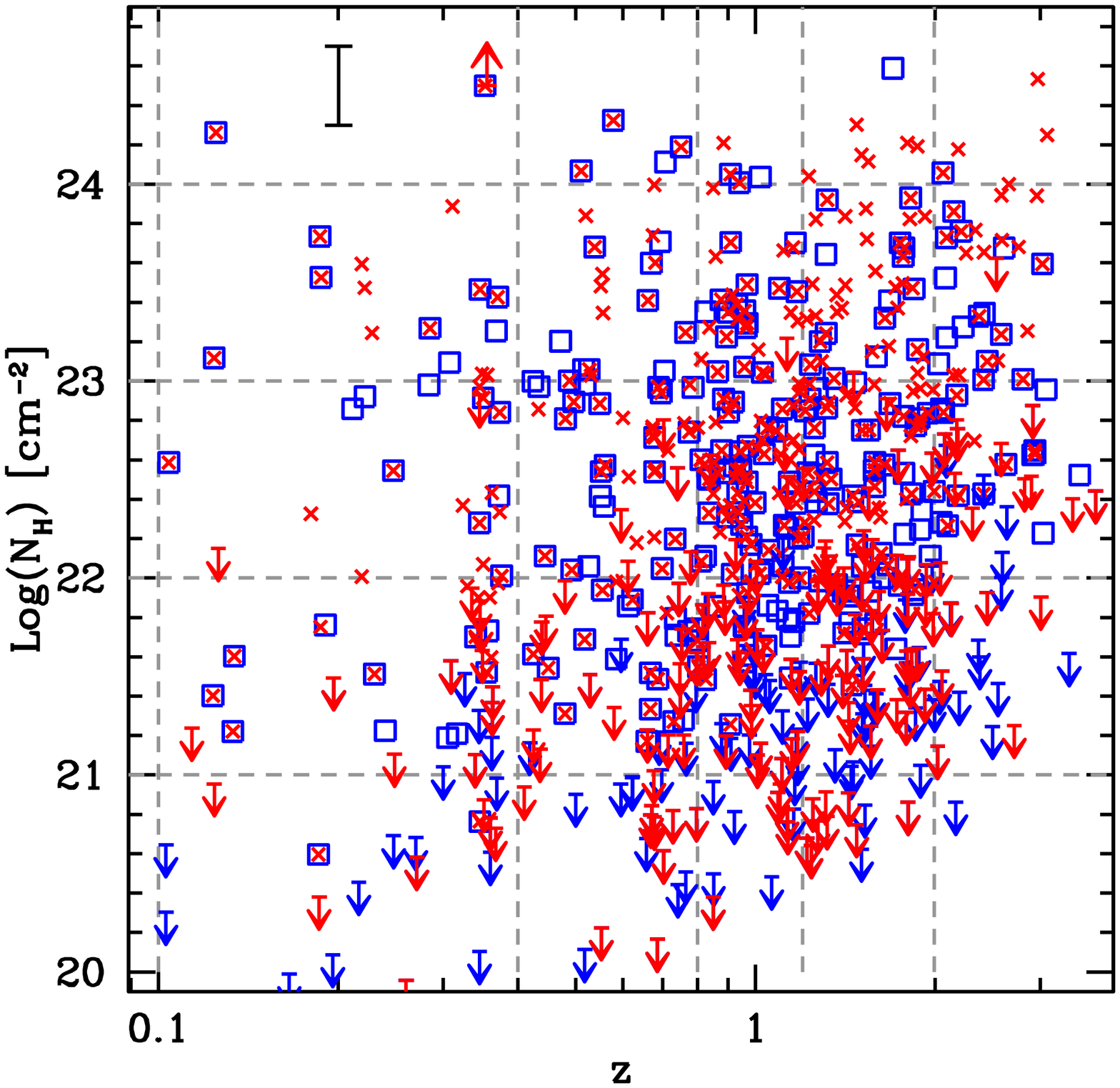}
 \caption{{\it Left}: Distribution of rest frame, absorption corrected 2-10 keV luminosity vs. redshift for the \xmm\ 
 (blue squares) and \chandra\ (red crosses) detected sources that are also \herschel\ detected (X-FIR sample). 
 The dotted (dashed) line marks the sensitivity limit of the \xmm\  (\chandra) surveys. 
 The redshift bins adopted in the text are marked by vertical gray dashed lines. 
 {\it Right}: Distribution of intrinsic column density vs. redshift for the sample. Symbols as in left panel. 
 Arrows show upper-limits for unobscured sources.
 The average $1\sigma$ error-bars on \lum\ and \nh\ are shown in the top left of both panels.
 }
 \label{fig:zlum}
 \end{center}
 \end{figure*}

We performed X-ray spectral fitting for all the \chandra\ and \xmm\ detected sources 
(from the catalogs of Brusa et al. 2010 and Civano et al. 2015 respectively)
with more than 30 counts, in  Marchesi et al. (2016) and Lanzuisi et al. (2013, 2015), respectively. 
This sample consist of 2333 individual sources 
(1949 \chandra\ and 1187 \xmm\ sources, with 803 source in common\footnote{For sources in common the \chandra\ data 
from Marchesi et al. (2016) are used, given the deeper coverage.}).

For all the \herschel\ detected sources in the COSMOS field 
(Lutz et al. 2011, $\sim17000$ with at least a detection at $>3\sigma$ in one of the FIR bands, from 100 to 500 $\mu m$), 
an SED deconvolution with 3 components --- stellar emission, AGN torus emission and SF-heated dust emission 
--- performed using photometric points from the UV to sub-mm, is available
from Delvecchio et al. (2014, 2015, D15 hereafter), following the recipe described in Berta et al. (2013).

We then selected all the \xmm\ and \chandra\ detected sources, having at least one FIR detection (and therefore SED 
deconvolution).
The final sample comprises 692 sources X-ray and FIR detected (the 'X-FIR sample' hereafter), 
all of them with an available redshift, 459 spectroscopic and 233 photometric (Civano et al. 2012, Brusa et al. 2010,
Salvato et al. 2009, Marchesi et al. 2015).
This is to date the largest sample of AGN/host systems for which X-ray spectral parameters,
such as column density and absorption-corrected 2-10 keV luminosity, are known in combination
with host properties such as \mstar\ and SFR.

\subsection{AGN properties}

Figure 1 (left) shows the distribution of \lum\ vs redshift for the X-FIR sample.
 The average $1\sigma$ error bar on \lum\ is shown in the upper left corner. 
The absorption-corrected \lum\ is affected by uncertainties related to both the 
number of net counts (observed flux uncertainties) 
and the spectral shape of each source (uncertainties on \nh\ and spectral slope).
Therefore, the errors have been derived, for each source, using the equivalent
in {\it Sherpa} (Fruscione et al. 2006) of the {\it cflux} model component in {\it Xspec} (Arnaud 1996),
applied to the best-fit unabsorbed powerlaw.
The flux and errors are then computed in the observed band corresponding to 2-10 keV rest-frame, 
and converted into luminosity.

  \begin{figure}[h!]
  \begin{center}
  \includegraphics[width=8.5cm]{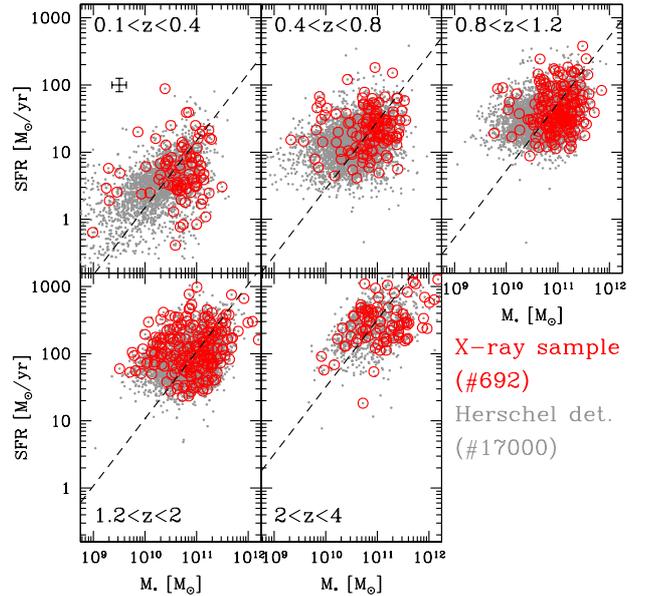}
  \caption{SFR vs. \mstar\ distribution for the entire sample of \herschel\ detected sources ($\sim17000$ sources, gray points) 
  and for the 692 sources with X-ray spectral analysis (X-FIR sample, red circles), divided in the five redshift bins defined in 
  Sec. 2.1. The dashed lines in each panel mark the redshift-dependent MS of Withaker et al. (2012).
   The average $1\sigma$ error-bars are shown in the top left as a black cross.
  }
  \label{fig:mstardet}
  \end{center}
  \end{figure}

The redshift bins that will be used in the following analysis 
are shown with vertical dashed lines. The intervals have been chosen with the aim of having a 
fairly large number of sources in each bin ($\sim80-160$) with reasonably narrow redshift interval.
The \lum\ bins that will be used in the following (1 bin per dex) are shown as horizontal dashed lines.

   \begin{figure*}[t]
   \begin{center}
   \includegraphics[width=6cm]{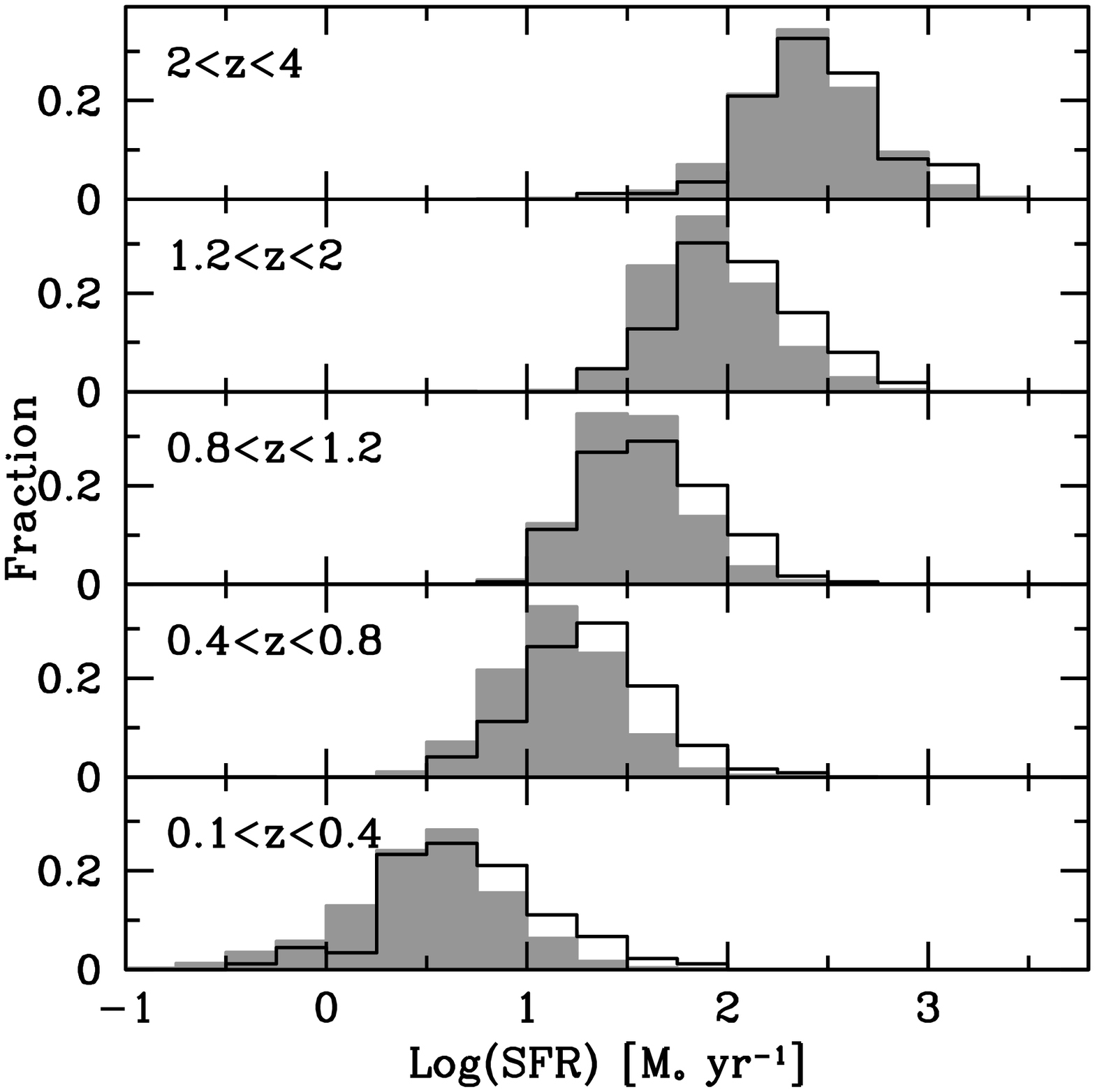}\hspace{1cm}\includegraphics[width=6cm]{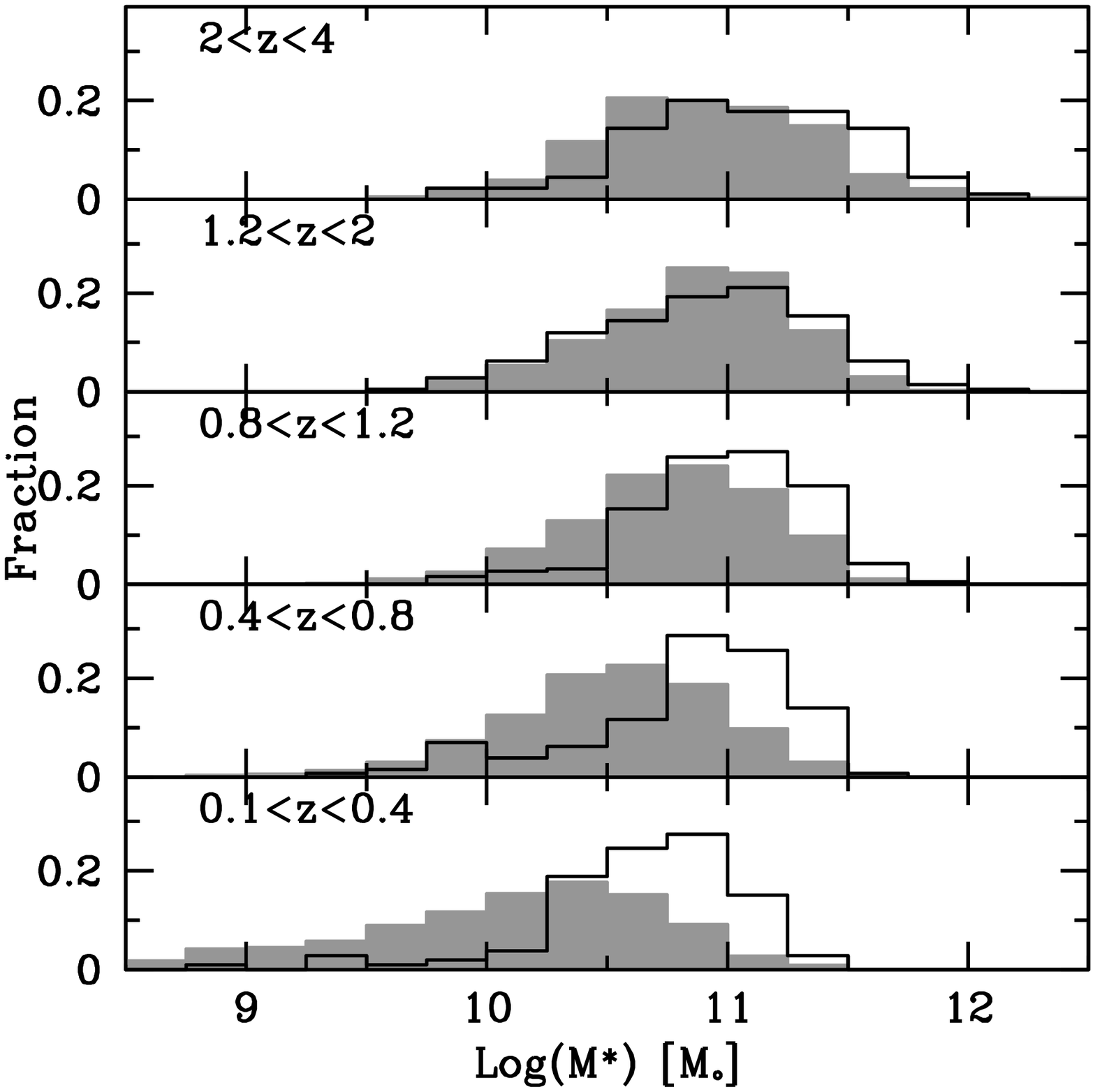}
   \vspace{1cm}
   \includegraphics[width=6cm]{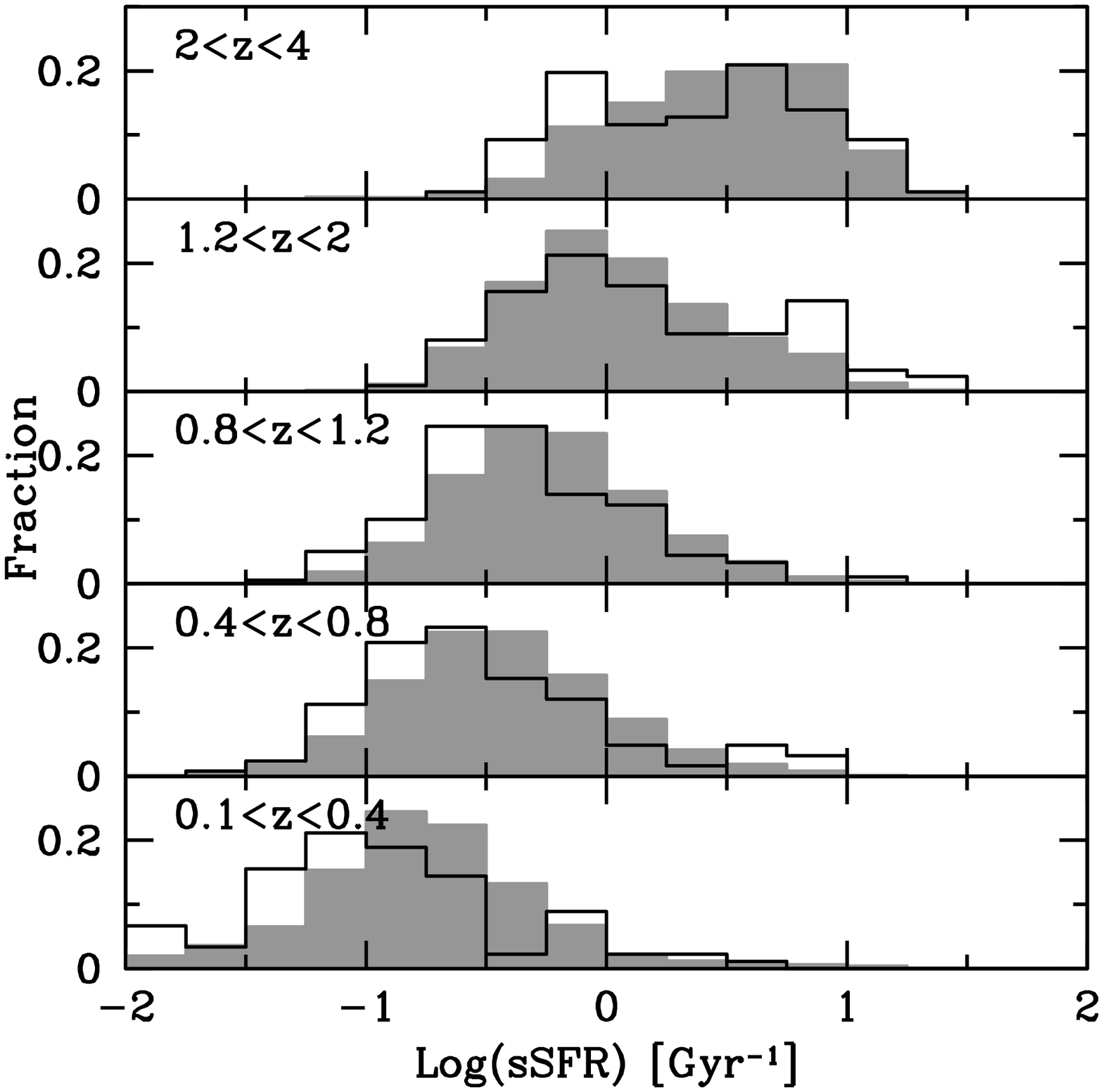}\hspace{1cm}\includegraphics[width=6cm]{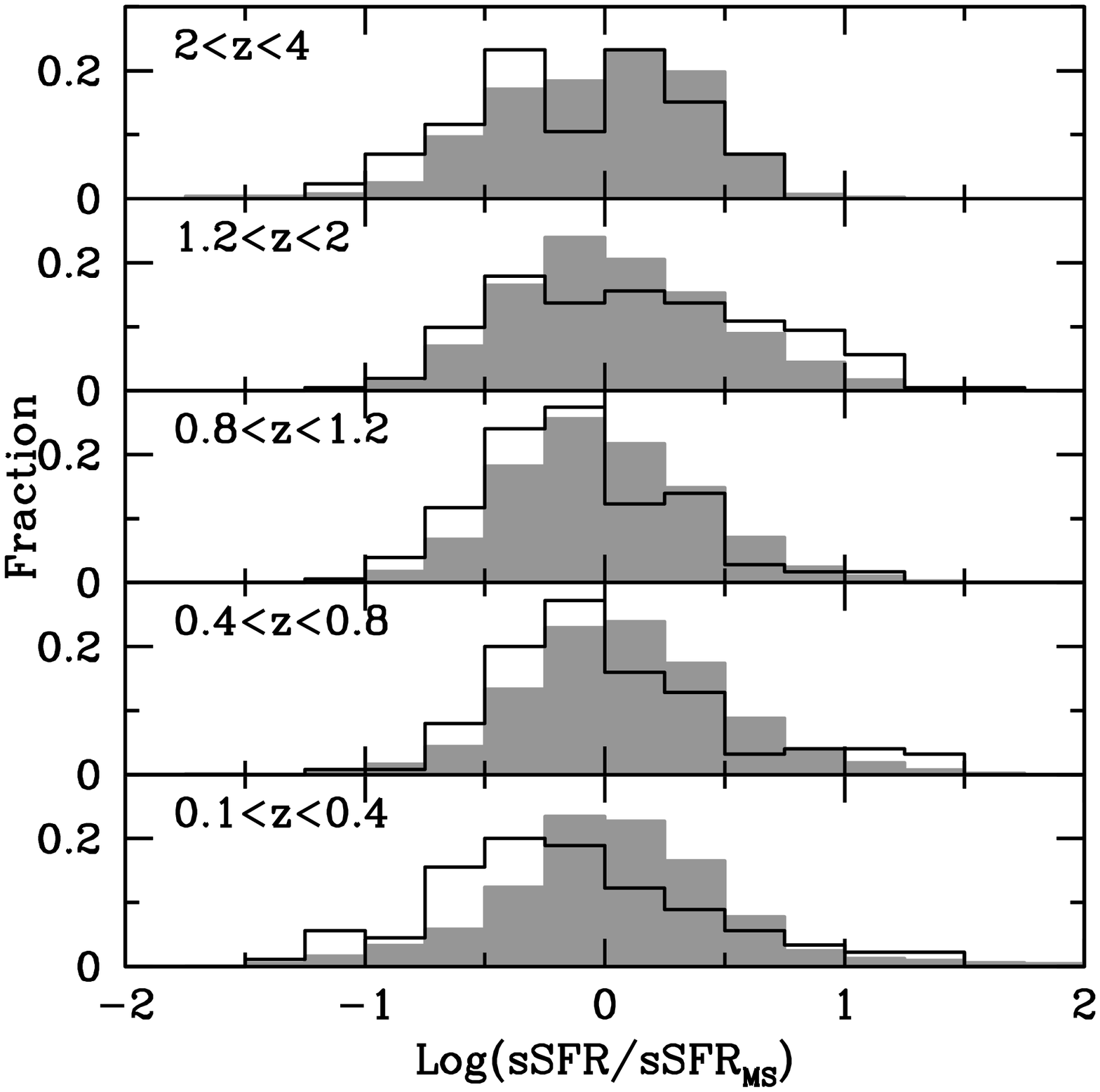} 
   \caption{Fractional distribution of the host properties, SFR, \mstar, sSFR, and MS offset, from top left to bottom right,  
   for the X-FIR sample (black open histogram), and for the whole \herschel\ sample (gray filled histogram), 
   in redshift bins. 
    }
   \label{fig:isto}
   \end{center}
   \end{figure*}

Figure 1 (right) shows the column density distribution for the X-FIR sample.
Arrows show sources for which the obscuration is constrained only by an upper-limit.
The average $1\sigma$ error bar on \nh\ is shown in the upper left corner.
The distribution of \nh\ from \xray\ spectral analysis has a clear upper boundary around CT column 
densities\footnote{
Lanzuisi et al. (2015a,b) present the CT sources detected by \xmm,
while Lanzuisi et al. 2017 in prep. will present the ones detected by \chandra.}
due to the strong flux decrement associated with CT obscuration in the 2-10 keV band. 
Also, the minimum measurable \nh\ increases with redshift, as the low energy cut-off due 
to obscuration move outside the observing band.

The global fraction of X-ray obscured sources (those with \nh$>10^{22} cm^{-2}$) in the X-FIR sample is $\sim50\%$, 
higher than the typical 
obscured fraction (30-40\%) of the X-ray samples in the \chandra- and \xmm-COSMOS 
(Lanzuisi et al. 2015, Marchesi et al. 2016).
Indeed, the FIR luminosity (and therefore \herschel\ detection rate) 
of type-2 AGN seems to be higher than for type-1 QSO (Chen et al. 2015).

\subsection{Host properties}

The host properties (SFR vs. \mstar) of the 692 sources in the X-FIR sample are shown in 
figure~\ref{fig:mstardet} (red circles) divided in five redshift bins as described above. 
The values are taken from D15: the SFR has been derived by converting the IR luminosity (rest $8-1000~\mu m$)
of the best-fitting galaxy SED 
(i.e. subtracting the AGN emission when present) with the SF law of Kennicutt (1998), 
scaled to a Chabrier (2003) initial mass function (IMF).
The \mstar\ is derived from the SED decomposition itself, and based on Bruzual \& Charlot (2003) models,
with a consistent IMF.
Table 1 (full version available on-line) summarizes the multi-wavelength properties of the sources in the X-FIR sample.

\begin{table*}
\begin{small}
\caption{Multi-wavelength properties of the 692 sources in the X-FIR sample.}
\begin{center}
\begin{tabular}{lccccccccccc}
\hline\\
ID  &  RA & DEC & z   & Log(\lirsf) & Log(\mstar) &   SFR  & Log(\nh)  & Log(\lum) & Log(\lbol) &  XID &  CID   \\
    & deg & deg &     & erg/s       &  \msun      & \msun/yr & cm$^{-2}$ & erg/s     &  erg/s     &      &           \\
(1) & (2) & (3) & (4) &  (5)        &   (6)       &   (7)    &  (8)      &  (9)      &    (10)    &  (11)&   (12)     \\      
\hline \\                                                                                                                                                                                                   
1846545 & 150.500 & 2.862 & 0.102s & $44.77\pm0.07$ & $10.20\pm0.13$ & 15.8 & $<20.77               $ & $41.05\pm0.27$ & 41.96 & 60095 & lid2100\\
1883498 & 150.065 & 2.929 & 0.102s & $43.69\pm0.16$ & $10.55\pm0.09$ &  1.3 & $<20.42               $ & $42.64\pm0.06$ & 43.8  & 5617  & lid385 \\
89570   & 150.372 & 1.609 & 0.104s & $44.03\pm0.08$ & $10.80\pm0.09$ &  2.8 & $22.58_{-0.03}^{+0.03}$ & $43.04\pm0.04$ & 44.29 & 2021  & cid1678\\
1612003 & 150.550 & 2.628 & 0.113s & $43.37\pm0.28$ & $10.58\pm0.09$ &  0.4 & $<21.36               $ & $41.63\pm0.36$ & 42.6  & ---   & lid3189\\
1197519 & 150.335 & 2.304 & 0.123s & $44.24\pm0.06$ & $10.58\pm0.09$ &  4.7 & $21.40_{-0.29}^{+0.62}$ & $41.2 \pm0.29$ & 42.11 & 1533  & cid967 \\
... & & & & & & & & & & & \\
\hline
\end{tabular}
\end{center} 
{\bf Notes.} Catalog entries are as follows: (1) Source ID from Capak et al. (2007); (2) and (3) right ascension and declination of the optical/IR counterpart;
(4) redshift ({\it s} for spectroscopic or {\it p} photometric); (5) Log(\lirsf) with $1\sigma$ errors; (6) Log(\mstar) with $1\sigma$ errors; 
(7) SFR derived from \lirsf; (8) Log(\nh) with $1\sigma$ errors or upper-limits; (9) Log(\lum) with $1\sigma$ errors; (10) Log(\lbol) computed from \lum\ using Marconi et al. (2004);
(11) and (12) XMM-COSMOS and Chandra-COSMOS IDs (from Brusa et al. 2010 and Marchesi et al. 2016 respectively).
The full table will be available in electronic form at the CDS via \url{http://cdsweb.u-strasbg.fr/cgi-bin/qcat?J/A+A/} .
\end{small}
\end{table*}

The host properties of the sample of \herschel\ detected sources (from D15, $\sim17000$ sources) 
are shown for comparison with gray dots. 
The average of the statistical $1\sigma$ error-bars resulting from the SED fit\footnote
{Systematic errors like, for example, uncertainties related to the adopted IMF or SF law, 
are not included in the error budget.}
are shown in the top left corner. 
The errors on \mstar\ follow a log-normal distribution, with average $\langle err($\mstar$) \rangle=0.14$ dex
and standard deviation $\sigma=0.09$.
The mean error on SFR is $\langle err($SFR$) \rangle=0.10$ dex, and standard deviation 
of $\sigma=0.07$ as for \lirsf\ (see sec. 3.1), since the SFR is derived from \lirsf\ adopting a Kennicutt (1998) law.
The redshift-dependent MS of star forming galaxies
as described in Whitaker et al. (2012) is also shown in each panel.
The FIR selected sources broadly follow the MS relation. However, 
the \herschel-based selection is sensitive to the most star forming systems,
introducing a cut in SFR that moves towards higher values with increasing redshift. 
(e.g. Rodighiero et al. 2011, D15). 

X-ray detected AGN are preferentially found at the highest \mstar, i.e. 
the fraction of X-ray detected sources increases as a function of \mstar, in the first three redshift bins at least.
This is a well known effect (Kauffmann et al. 2003, Bundy et al. 2008, 
Brusa et al. 2009, Silverman et al. 2009, Mainieri et al. 2011, Santini et al. 2012, Delvecchio et al. 2014). 
Aird et al. (2012) suggested that it is the result of an observational bias, such that 
more massive galaxies (i.e. more massive BHs),
can be detected, at a given X-ray flux limit, with a variety of accretion rates, while lower mass systems 
can be detected only if they have a high accretion rate.
This, combined with a steep Eddington ratio distribution (i.e. sources with low Eddington ratio are
much more common than sources with high Eddington ratio) can explain the observed 
\mstar\ distribution (see also Bongiorno et al. 2012).

In our case there is a threshold at around $\log$\mstar$\sim$10.5 \msun\ in the first 3 redshift bins.
A simple calculation shows that this value can be roughly derived from the X-ray flux limit of the \chandra\
and \xmm\ surveys, using standard values for bolometric corrections ($k_{Bol}=10-30$), Eddington ratios (\edd$\sim0.05$)
and BH-host mass ratios (\mstar$/M_{BH}=1000-3000$). 
A more detailed study of the Eddington ratio distribution that can be derived from the \mstar\ and \lum\ distributions,
will be presented in Suh et al. (2017 submitted).

Several studies in the local Universe suggest that the fraction of galaxies hosting an AGN
increases also with IR luminosity (e.g. Lutz et al. 1998, Imanishi et al. 2010, Alsonso-Herrero et al. 2012, Pozzi et al. 2012).
We also tested that the observed threshold in mass is not driven by our requirement of \herschel\ detection:
also using the \mstar-SFR distribution of Bongiorno et al. 2012, computed for the full XMM-COSMOS catalog,
a drop in the number of X-ray detected AGN below $\log$\mstar=10.2-10.4 \msun\ is visible up to $z=2.5$.

The consequence of this selection effect is that
the X-FIR sample has a \mstar\ distribution shifted toward
higher \mstar\ with respect to the global \herschel\ sample (Fig.~\ref{fig:isto} top right).
The distribution of SFR for the X-FIR sample, instead, is roughly consistent 
with that of the global \herschel\ sample (Fig.~\ref{fig:isto} top left).
This have important implications when measuring, e.g., sSFR and MS offset
(Fig.~\ref{fig:isto} bottom left and right): due to this selection effect
the X-FIR sample has lower sSFR with respect to the MS of star forming galaxies (or to the \herschel\ sample), 
if the two samples are not properly mass-matched (Silverman et al. 2009, Xue et al. 2010).

 \begin{figure}[h]
 \begin{center}
 \includegraphics[width=8cm]{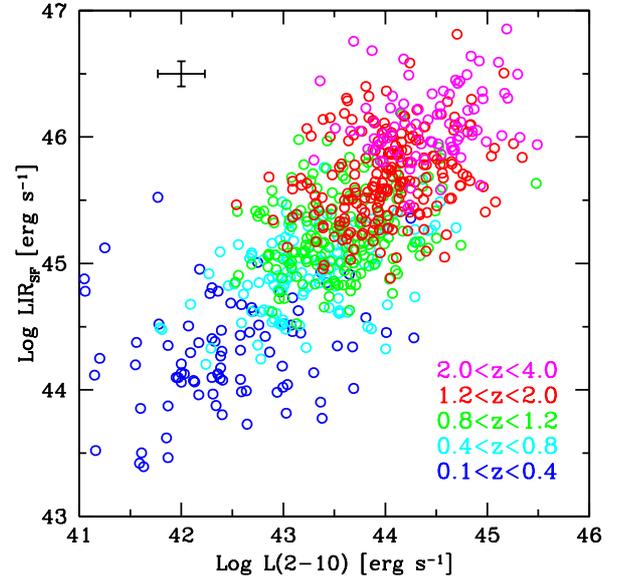}
 \caption{\lum\ vs. \lirsf\ for the X-FIR sample. Different colors represent different redshift bins: 
 blue for $0.1<z<0.4$, cyan for $0.4<z<0.8$, green for $0.8<z<1.2$, red for $1.2<z<2$ and magenta for $2<z<4$. 
 The average $1\sigma$ errors on \lum\ and \lirsf\ are shown in the upper left corner.
 }
 \label{fig:lumlum1}
 \end{center}
 \end{figure}

\section{\lum\ vs. \lir\ distributions}

\subsection{Partial correlation analysis}

The two quantities that have been more often used in order to look for BHAR-SFR correlations
are the AGN luminosity, often represented by the \lum, and the SF luminosity
in the form of \lir\ (or L$_{60 \mu m}$, Santini et al. 2012, Rosario et al. 2012, Chen et al. 2013).
It is generally assumed that the total FIR luminosity is not significantly affected by any contamination 
from the AGN emission. However, recent studies have shown that the AGN 
may contribute significantly to the IR emission and in some case even in the FIR band (Symeonidis et al. 2016). 
Therefore, the SFR derived directly from FIR photometry can be overestimated, especially in high luminosity AGN hosts.
Thanks to the SED decomposition available, we will use in the following the \lir\, computed for the SF 
component only (\lirsf\ hereafter), after subtracting the AGN contribution, modeled with the SED templates 
of Fritz et al. (2006, see also Feltre et al. 2012).
This will allow us to avoid introducing a spurious correlation between AGN and SF luminosity, especially at the
highest luminosities. 

Clearly two luminosities are always correlated in any sample that is flux limited in both directions, 
due to the combination of the luminosity-distance effect
and of the fact that the sources tend to cluster at the flux limit (Malmquist bias, e.g. Feigelson \& Berg 1983).
Figure~\ref{fig:lumlum1} shows the distribution of \lum\ vs. \lir\ for the X-FIR sample.

 The $1\sigma$ errors on \lirsf\ follow a log-normal distribution with average value 
$\langle err($\lirsf$)\rangle=0.10$ dex, and standard deviation of $\sigma=0.07$. 
As mentioned in sec. 2.1, the errors on the absorption corrected luminosity 
follow a much broader distribution, depending both on the number
of counts available and on the spectral shape. 
They range from $\simlt0.1-0.2$ dex for bright, unobscured
sources, to $\sim0.5-1.0$ dex for faint and highly obscured sources.
The average value of the $1\sigma$ error is $\langle err($\lum$)\rangle \sim0.23$ dex, 
with standard deviation $\sigma=0.18$.  
We show the average errors with a black cross in the left panel of Fig. 4, while 
the specific value for each source is used in the following analysis.

 \begin{figure*}[t]
 \begin{center}
 \includegraphics[width=8cm]{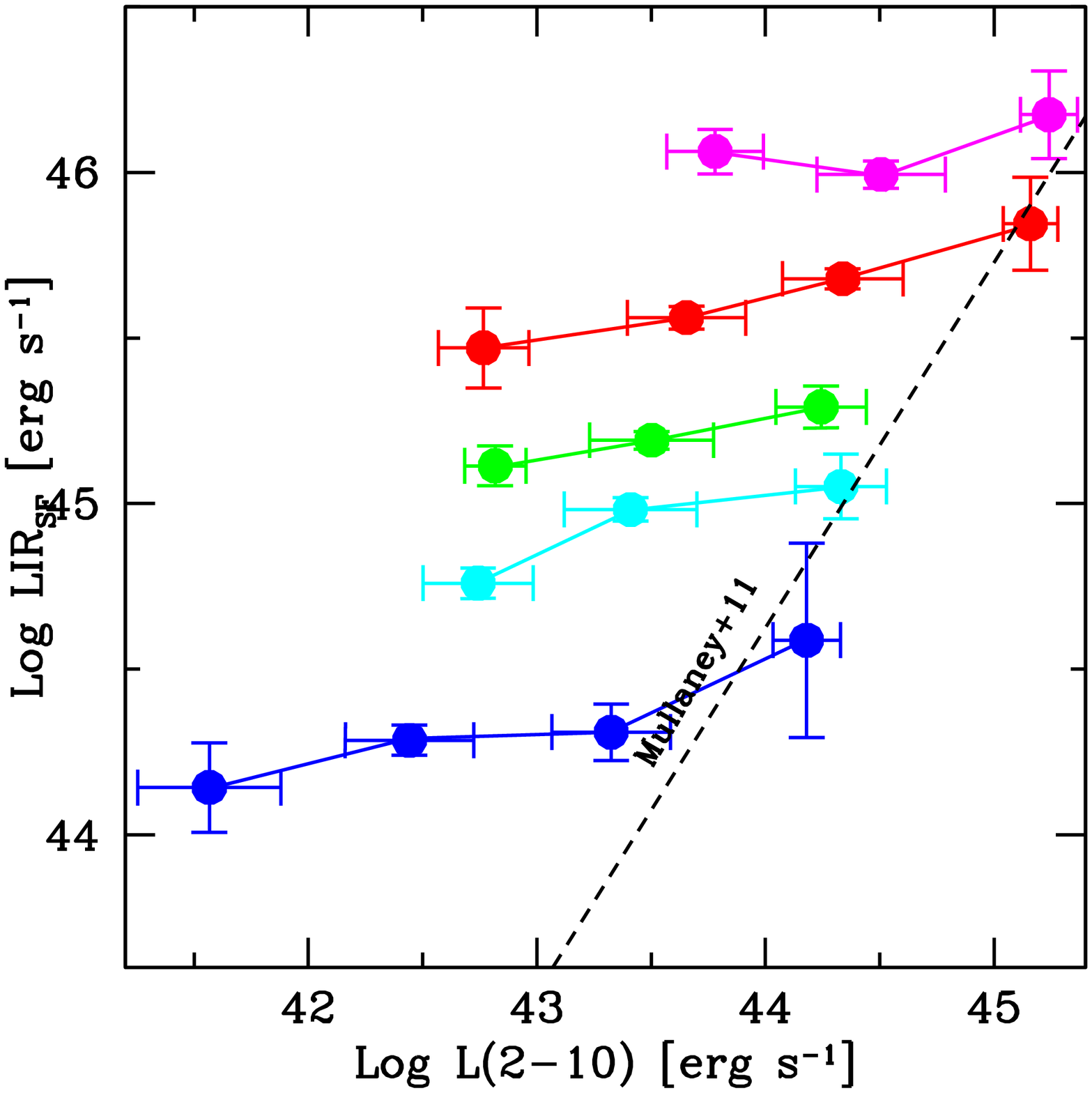}\hspace{1cm}\includegraphics[width=8cm]{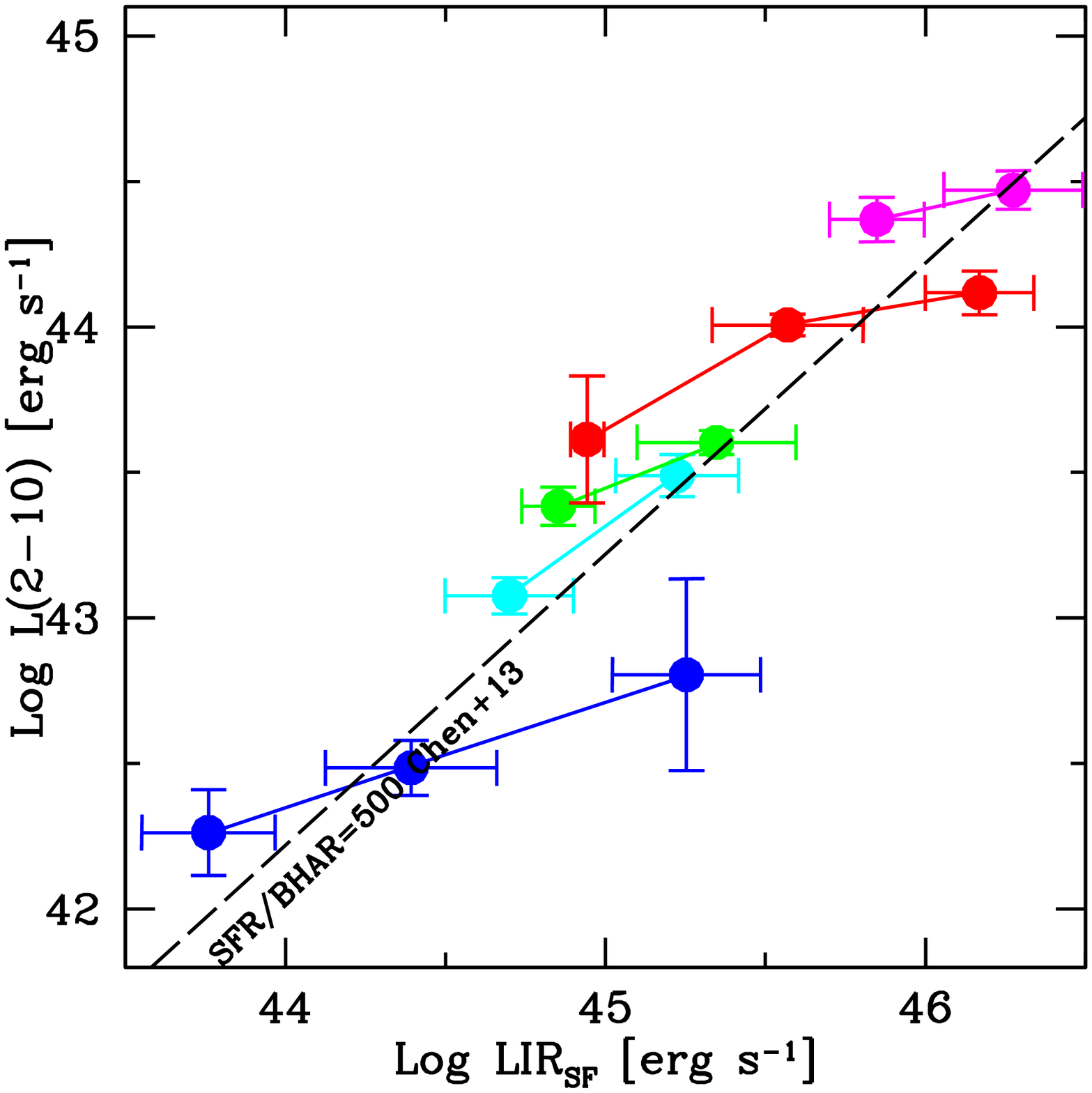}
 \caption{{\it Left}: Average Log(\lirsf) in bins of Log(\lum) in five redshift bins.
 The short dashed line is the correlation derived in  Mullaney et al. (2011) for a pure AGN SED.
  {\it Right}: Average Log(\lum) in bins of Log(\lirsf). The long dashed line represents
 a constant SFR/BHAR of 500, from C13. 
  In both panels the vertical error-bars are computed through a bootstrap re-sampling procedure, while
the horizontal error-bars show the $1\sigma$ dispersion of that bin.} 
 \label{fig:lumlum2}
 \end{center}
 \end{figure*}

In order to look for intrinsic correlations between these two quantities,
one possibility is to compute the {\it partial} Spearman rank correlation between two variables
in presence of a third, and to assess the statistical significance of such correlation (e.g. Macklin 1982).
To derive the correlation coefficient between \lum\ and \lirsf, conditioned by the distance, 
$\rho$(\lum, \lirsf, $\dot{z}$), we evaluate the
Spearman coefficient $\rho$ related to each couple of parameters and then combined them according to the expression:
\begin{equation}
\rho(a,b,\dot{c})=\frac{\rho_{ab}-\rho_{ca}\rho_{bc}}{\sqrt{(1-\rho_{ca}^2)(1-\rho_{bc}^2)}}
\end{equation}
(Conover 1980) which returns the partial correlation between $a$ and $b$, corrected for the dependency on $c$. 
The resulting $\rho$ is 0.15, and the associated confidence level, in terms of standard deviations,
that the first two variables are correlated, independently of the influence of the third,
is $\sim3.7\sigma$, following eq. 6 of Macklin (1982).
Therefore, the two quantities appear to be significantly correlated, after the effect of redshift on both of them
is taken into account.

\subsection{Redshift bins}

The second approach, often used in the literature, is to define as narrow as possible redshift bins,
to minimize the distance effect, and look for correlations between the two quantities.
Thanks to the large sample collected in this work, we can divide the sample in five redshift bins.
For every redshift bin, a large distribution in both luminosities can be observed, with the typical
luminosity increasing with redshift (Fig.~\ref{fig:lumlum1}).

Most of the observational works mentioned in Sec.~\ref{sec:intro} looked for
the distribution of {\it average} \lirsf\ in \lum\ bins, or  {\it average} \lum\ in \lirsf\ bins 
 (but see Gruppioni et al. 2016).
Both hydrodynamical simulations (e.g. Volonteri et al. 2015a,b) and semi-analytic model (e.g. Neistein \& Netzer 2014) 
show that, in the \lirsf-\lum\ plane, there may be the superimposition of a weak correlation for the bulk of the population, 
and a strong correlation only for the most extreme merger phases, corresponding to the highest \lum\ and \lirsf.
If the underlying distribution shows such a complex shape, the results of the two approaches (average \lirsf\ in \lum\ bins
or average \lum\ in \lirsf\ bins) may be very different.

In Fig.~\ref{fig:lumlum2} left, we show the result of plotting {\it average} \lirsf\ in bin of \lum\ (both in log scale), 
in five redshift bins. 
As can be seen, there is no correlation at all \lirsf\  and, as expected, there are no sources below the relation 
computed for a pure AGN template in Mullaney et al. (2011),  similar to the one of Netzer et al. (2009). 
Following this approach, we are therefore able to reproduce the results of Shao et al. (2010), Rosario et al. (2012)
and others, that claim no correlation between AGN activity and SF over several orders of magnitudes in luminosity.

On the other hand, computing {\it average} \lum\ in \lirsf\ bins (in log scale), from the same bivariate distribution, 
gives different results. Fig.~\ref{fig:lumlum2}, right, shows that, at all redshifts, 
the average \lum\ correlates with the \lirsf\ and the binned points are close to the 
SFR/BHAR$\sim500$ ratio found in Chen et al. (2013, C13 hereafter).

In both panels, we computed the error on the average \lum\ and \lirsf\ through
a bootstrap re-sampling procedure, as done in several previous works. 
For each bin with N sources,
we randomly extract N sources, allowing repetitions, and computed the mean value. The process is iterated $10^4$ times,
and the standard deviation of the mean is taken as error on the average SFR.

\begin{table*}
\caption{Slopes $\alpha$ and intercept $\beta$ of the linear LS fit of (\lirsf | \lum), 
(\lum | \lirsf), and of the bisector estimator, in each redshift bin.
The first set of slopes and intercepts refers to a relation in the form
$Log~$\lirsf $=45+\alpha\times(Log~$\lum $ -44)+\beta$, while the second and third
in the form $Log~$\lum $=44+\alpha\times(Log~$\lirsf $ -45)+\beta$.
}
\begin{center}
\begin{tabular}{lccccccc}
\hline\\
z-bin & & \multicolumn{2}{c}{LS (\lirsf\ | \lum)} &  \multicolumn{2}{c}{LS (\lum\ | \lirsf)}  & \multicolumn{2}{c}{bisector(\lum\ , \lirsf)} \\
\hline                                 
                &  & $\alpha$        & $\beta$         & $\alpha$        &   $\beta$      &  $\alpha$         &   $\beta$ \\
$0.1\leq z<0.4$ &  & $0.07\pm0.06$   & $-0.61\pm0.09$  & $ 0.44\pm0.13 $ & $-1.24\pm0.21$ &  $ 1.28\pm0.45$ & $-0.65\pm0.20$ \\
$0.4\leq z<0.8$ &  & $0.20\pm0.10$   & $0.06\pm0.04$   & $ 0.80\pm0.17 $ & $-0.68\pm0.05$ &  $ 1.21\pm0.34$ & $-0.67\pm0.11$ \\
$0.8\leq z<1.2$ &  & $0.12\pm0.09$   & $0.25\pm0.03$   & $ 0.61\pm0.15 $ & $-0.56\pm0.04$ &  $ 1.30\pm0.25$ & $-0.75\pm0.09$ \\
$1.2\leq z<2.0$ &  & $0.16\pm0.09$   & $0.62\pm0.04$   & $ 0.48\pm0.12 $ & $-0.34\pm0.08$ &  $ 1.29\pm0.15$ & $-0.86\pm0.15$ \\
$2.0\leq z<4.0$ &  & $0.01\pm0.08$   & $1.02\pm0.04$   & $ 0.29\pm0.18 $ & $0.10\pm0.20$  &  $ 1.16\pm0.57$ & $-0.85\pm0.17$ \\
\hline\\
\end{tabular}
\end{center} 
\end{table*}

The two approaches described above are the equivalent of computing the forward and 
inverse linear regression of one variable over the other. 
Table 2 reports the slopes $\alpha$, intercept $\beta$ and associated error,
for each redshift bin, in the log-log space, of the least square (LS) fit\footnote{ 
The LS fit is performed with the BCES code (Akritas \& Bershady 1996), adopting $10^4$ bootstrap re-samplings.
Similar results are obtained using the LINMIX code (Kelly et al. 2007).}
of \lirsf\ as a function of \lum\ (hereafter \lirsf\ | \lum\ ), and \lum\ as a function of \lirsf\ (hereafter \lum\ | \lirsf\ ),
respectively\footnote{
In the first case slopes and intercepts refer to a relation in the form
$Log~$\lirsf $=45+\alpha\times(Log~$\lum $ -44)+\beta$, while in the second
in the form $Log~$\lum $=44+\alpha\times(Log~$\lirsf $ -45)+\beta$.}.
Indeed the slopes in the left panel are all consistent with 0 within $\sim2\sigma$ c.l.. On the other hand,
LS fits of (\lum\ | \lirsf\ ) give steeper correlations at all z bins, and slopes not consistent with 0 at $\sim3\sigma$ c.l.

The SFR/BHAR$\sim500$ ratio plotted in Fig. 5 is the one found in C13, 
for a sample of 121 FIR selected AGN-hosts at $0.25<z<0.8$.
To compare with their results we should look at our first two z-bins: While the z-bin    
$0.1\leq z<0.4$ has a very flat (\lum\ |\lirsf\ ) slope, possibly due to 
the small volume sampled, the $0.4\leq z<0.8$ interval 
shows a correlation with slope consistent with 1 at $\sim1\sigma$, 
therefore in broad agreement with the C13 findings. 
Interestingly, we can extend up to $0.8\leq z<1.2$ the redshift range for which a correlation roughly consistent with
SFR/BHAR$\sim500$ can be found. Above this redshift interval, the slopes become flatter.
Therefore, we found a strong (almost linear) correlation between $\log$\lum\ and $\log$\lirsf,
for (\lum\ | \lirsf\ ) at redshifts lower than the peak 
of the SF and AGN activity, i.e. between 4 and 8 Gyr ago, while at higher redshift the correlation
is still present but weaker.

The exact value of the ratio SFR/BHAR in terms of \lum\ and \lirsf\ 
depends strongly on the assumptions made to scale between these quantities, 
i.e. the accretion efficiency and bolometric correction 
in the first case, and the SF law and initial mass function (IMF) in the second.
C13 derived the SFR from \lirsf\  using the Kennicutt (1998) relation, 
modified for a Chabrier IMF (Chabrier 2003) and the BHAR from \lum\
using a constant $k_{bol}=22.4$ and accretion efficiency of 0.1.  
They use as reference the value of SFR/BHAR$\sim500$ derived from the M$_{\rm Bulge}$/M$_{\rm BH}$ 
ratio observed in Marconi et al. 2004. 
The authors suggest that the fact that the detected sources sit on the SFR/BHAR$\sim500$ ratio is a coincidence,
due to the ratio between the X-ray and FIR flux limits in the Bo\"otes field.

In the X-FIR sample in COSMOS we have a factor of $\sim10$ deeper X-ray data 
(taking into account the flux limit corresponding to our spectral analysis requirements), 
while the \herschel\ data are only a factor 2-3 deeper ($\sim8$ mJy at $110\mu m$ and 8 mJy at $250\mu m$) than in Bo\"otes.
Nonetheless, our X-FIR detected sample sit close to the C13 relation.
We underline that, in both cases, the X-ray and FIR detected sources are a small minority of both the original X-ray
and FIR samples (few \% up to $\sim20$\%), and as discussed also in C13, the flux limit has an important role 
in the observed properties of the detected sources alone.

 \begin{figure}
 \begin{center}
 \includegraphics[width=8cm]{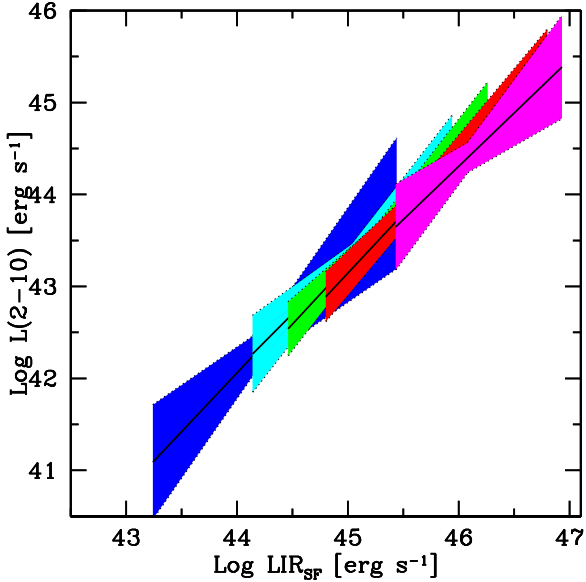}
 \caption{Linear regression for \lum\ and \lirsf\ computed for each redshift bin with the bisector 
 estimator in BCES.  The color code is the same as Fig. 4. 
}
 \label{fig:lumlum3}
 \end{center}
 \end{figure}

As discussed in Hickox et al. (2014) and Volonteri et al. (2015), a 
possible physical explanation for this behavior is that, 
when looking at left panel of fig. 5, we are averaging a slowly 
changing quantity, such as the host SFR, of galaxies grouped on the basis of the rapidly changing AGN \lum.
In the right panel, instead, the average \lum\ of a large sample of sources grouped on the basis of the slowly 
changing SFR, allows us to recover the underlying, long term correlation between AGN activity and SFR.
In the same way, from a statistical point of view, it may be reasonable to interpret the \lum\ as the dependent variable, 
in this context, as it has larger uncertainties with respect to \lirsf\ (Hogg et al. 2010), 
both in terms of  measurement errors (see sec. 3.1) and {\it noise} (i.e. variability).

If we instead assume that in this case there is no ``dependent'' and ``independent'' variables 
(see e.g. Tremaine et al. 2002, Novak et al. 2006),
the two variables may need to be treated symmetrically.
We used again the BCES code, to derive slope and intercept,
and their standard deviation, using a symmetric estimator such as the bisector regression\footnote{We 
recall that the BCES estimators, both the LS and the symmetric ones, are not immune from biases
that arise from data truncation, which is the case for flux-limited samples (see  Akritas \& Bershady 1996).}
(Isobe et al. 1990). 
The results are shown in figure~\ref{fig:lumlum3}, while the slopes and intercepts are 
reported in Table 2. At all redshift bins, the slopes of the linear regression,
although always larger than 1, are consistent with 1 within $1\sigma$ c.l.

\subsection{Effect of Contamination}

Since the \herschel\ (Pilbratt et al. 2010) PACS and SPIRE point spread functions
are much larger than the one in the optical and NIR bands,
going from $\sim5\arcsec$ to $\sim36\arcsec$ FWHM (Poglitsch et al. 2010; Griffin et al. 2010), 
there is the possibility that the FIR flux of our sources
is contaminated by unresolved neighbors (see e.g. Scudder et al. 2016). 

We verified the effect of contamination by excluding from the X-FIR sample all sources with a 
second HST catalog entry, from the ACS F814W (I-band) catalog (28.6 AB limiting magnitude, Scoville et al. 2007, Koekemoer et al. 2007).
We choose a circular area of diameter $8\arcsec$ around the optical position.
While this distance is not enough to ensure negligible contamination, it has been 
chosen in order to retain a sufficient number of sources to allow an analysis in all the five redshift bins.
The 146 ``isolated'' sources obtained in this way show the same behavior described above, with a flat 
distribution of {\it average} \lirsf\ computed in bin of \lum, and an almost linear correlation of {\it average} \lum\
computed in bin of \lirsf.

We also verified that sources with a single PACS or SPIRE detections (more subject to contamination)
do not affect our results. Indeed, excluding the 154 (out of 692) sources with only one detection (at $3\sigma$) either in
PACS or SPIRE photometry, does not change the results presented in sec. 3.2 and in the following paragraphs.

\subsection{\lbol, \mstar, sSFR and MS offset}

 \begin{figure}[t]
 \begin{center}
 \includegraphics[width=8cm]{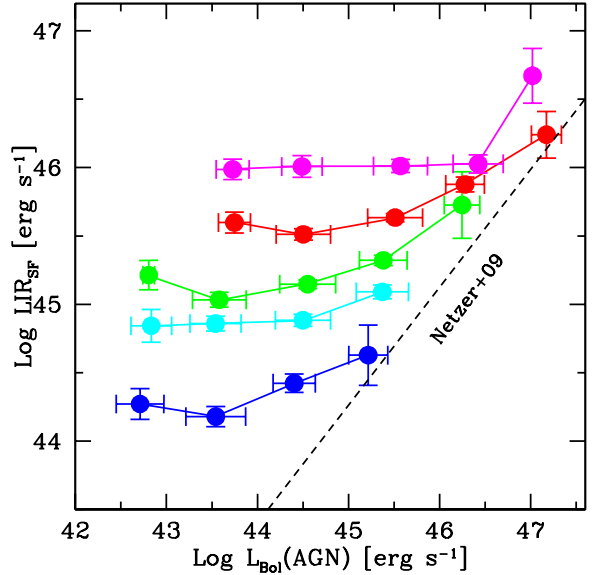}
 \caption{Average Log(\lirsf) in bin of \lbol, for the X-FIR sample. 
 The dashed line is the relation found in Netzer et al. (2009)
 for AGN dominated systems. Error-bars computed as in Fig. 5.
 }
 \label{fig:lirlbol}
 \end{center}
 \end{figure}

 \begin{figure*}[t]
 \begin{center}
 \includegraphics[width=8cm]{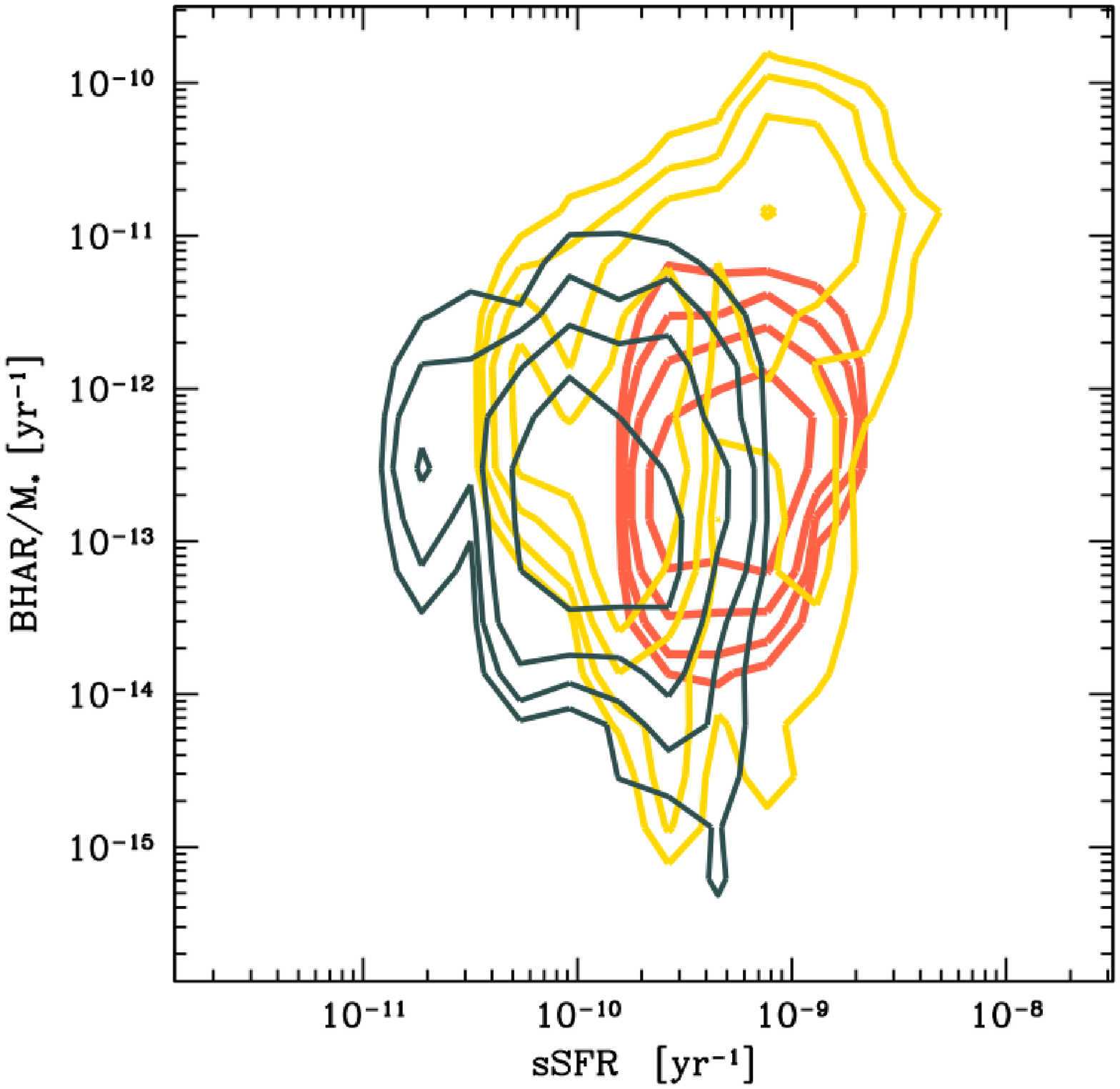}\hspace{0.5cm}\includegraphics[width=8cm]{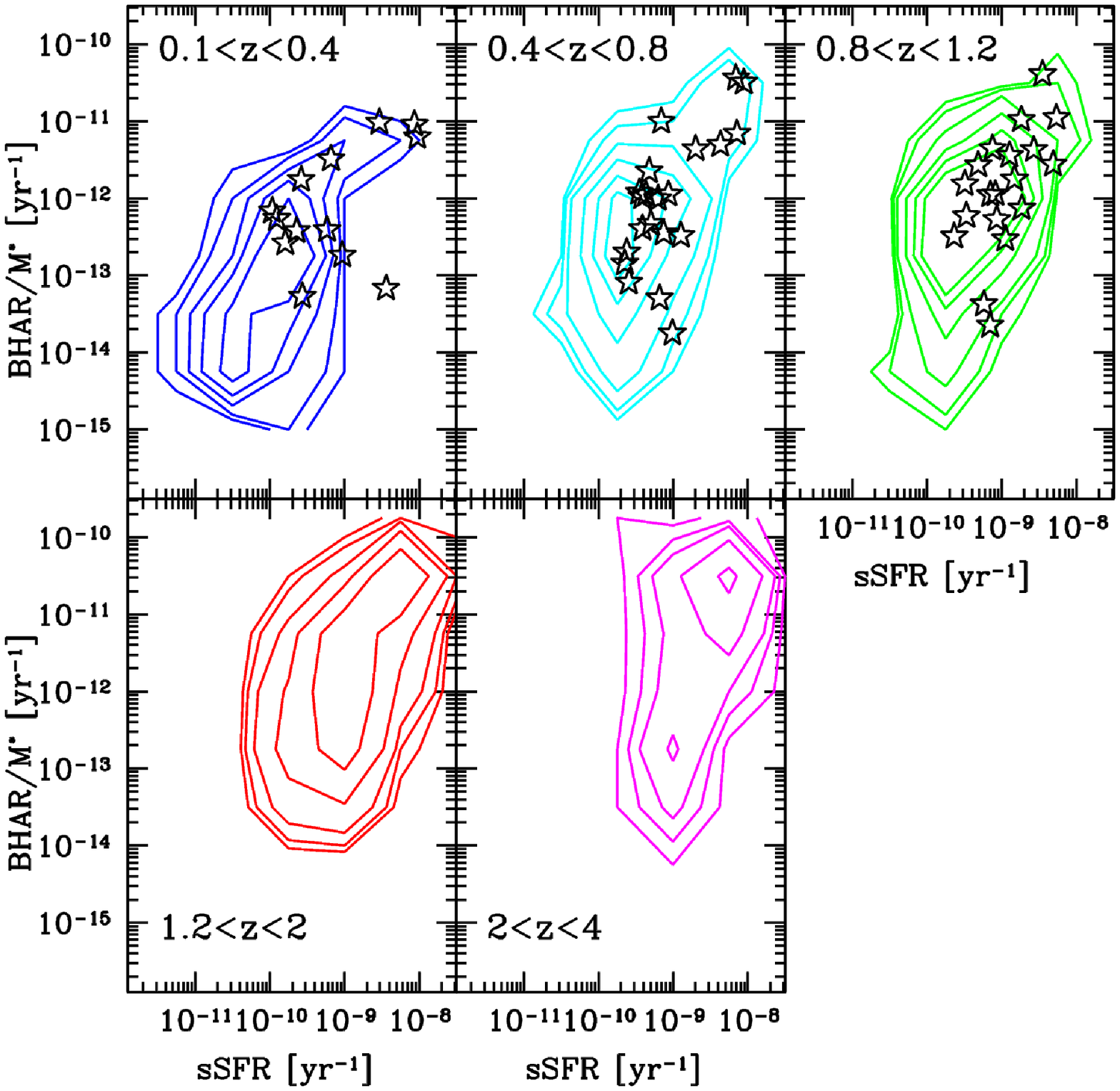}
 \caption{{\it Left}: BHAR/\mstar\ rate vs. sSFR contours obtained from the simulations presented in 
  Volonteri et al. (2015a). In red the stochastic phase, in yellow the merger phase and in 
  black the remnant phase.
  {\it Right}: BHAR/\mstar\ vs. sSFR contours observed in COSMOS 
  in five redshift bins. Sources that are in a major merger state in the first three redshift bins
  are marked with black stars.}
 \label{fig:simul}
 \end{center}
 \end{figure*}

Several authors have used the AGN bolometric luminosity (\lbol), instead of the \lum, 
to look for correlation with the \lir\ or SFR.
The \lbol\ is generally derived from the \lum\ through a luminosity dependent bolometric correction (e.g. Marconi et al. 2004, 
Lusso et al. 2012).
The net effect of this procedure is to stretch the horizontal axis of Fig. 5, left (the high \lum\ sources have a higher 
X-ray bolometric correction than the low \lum\ ones), 
while keeping the \lirsf\ fixed.
In Fig.~\ref{fig:lirlbol} we show the result of this approach (here we used the Marconi et al. 2004
luminosity dependent bolometric correction, but the Lusso et al. 2012 relation would have the same effect): 
in each redshift bin, the sources populating the highest \lum\ bin are now spread in two \lbol\ bins
and the last \lbol\ bin at each redshift is now populated by a smaller number of more extreme sources.
The relation found locally for AGN-dominated systems in Netzer et al. (2009) is also shown.
Once again, we are able to reproduce results obtained in other works (Shao et al. 2010, Rosario et al. 2012).
However, we are now confident that this result is not in disagreement with what shown in Fig. 5 (right),
and the apparent contradiction is only dependent on the way the data are analyzed and grouped,
as shown in e.g. Volonteri et al. (2015) and Dai et al. (2015).

Finally, we found a flat distribution when computing {\it average} \lum\ in bins of \mstar, sSFR and MS offset,
and {\it average} \mstar, sSFR and MS offset, 
in bins of \lum, in all the five redshift bins.   
Indeed, no significant partial correlation is found, between any pair of these quantities,
following the approach described in sec. 3.1 to take into account the redshift effect, that affects 
also \mstar, SFR and sSFR ($\sigma<<1$ in all cases).
We stress however that the range of \mstar\ covered by our sample is limited to the very 
high mass end, $10<Log($\mstar$)<12$  (to be compared with the underlining galaxies \mstar\ distribution shown
e.g. in Laigle et al. 2016, $7<$Log(\mstar)$<12$  in the same redshift interval). 
Deeper X-ray surveys are needed to investigate the dependency of \lum\ 
with this crucial quantity.

\section{Comparison with simulations}

Here we compare our results with predictions from the simulations of galaxy mergers presented in Volonteri et al. (2015a).
They are based on very high spatial and temporal resolution simulations, covering a 
large range of initial mass ratios (1:1 to 1:10), several orbital configurations, and gas fraction (defined as $M_{gas}/$\mstar)
in the range $f_{gas}=0.3-0.6$.
The very high resolution imposes a limit on the mass of the simulated galaxies, that typically 
have \mstar$\sim(2-8)\times10^{9}$ \msun, i.e. much smaller than the typical mass of our observed galaxies (see Fig. 3).
The process is divided into three phases: the {\it stochastic} phase, in which the galaxies behave as they do in isolation,
that lasts until the second pericenter; the  {\it merger} phase characterized by strong dynamical torques
and angular momentum loss; the {\it remnant} phase, that starts when the angular momentum returns to be constant in time.
While the stochastic and remnant phases have the same duration (by construction), the merger phase is much shorter (typically 1/10
of the total).

To compare our data with this set of simulated galaxies, we converted the  AGN bolometric luminosity
into a BH mass accretion rate (BHAR), by assuming an efficiency of $\eta=0.1$ 
(e.g. Fabian \& Iwasawa 1999) and dividing it by the host stellar mass, to obtain a specific BHAR (sBHAR) relative to 
the host mass, rather than to the BH mass.  We choose to do so because, from an observational point of view, 
the determination of the \mstar\ (from SED fitting) is much less uncertain 
(see sec. 2.2 for the error budget in our sample) than that of \mbh, and is available 
for both type-1 and type-2 AGN. This value is then compared with the sSFR for each source.
The contours of global (within 5 kpc) sSFR vs. sBHAR, obtained from the simulations
for the three different phases (stochastic, merger and remnant), are color coded in Fig.~\ref{fig:simul} 
(left) with red, yellow and black, respectively. 

The results from the X-FIR sample are shown in Fig.~\ref{fig:simul} (right)
for the five redshift bins. As can be seen the observed contours in the low redshift bins
span a similar range of physical properties, with respect to simulations, 
with the bulk of the population concentrated between $5\times10^{-11}$ and $5\times10^{-9}$ yr$^{-1}$
in sSFR, and between $10^{-14}$ and $10^{-11}$  yr$^{-1}$ in sBHAR,
and with a tail at higher sSFR and sBHAR, possibly produced by sources in the merger phase as in the simulations 
(yellow contours). Interestingly, the importance of this tail grows with increasing redshift, 
even if the selection effect in both directions
must be taken into account.

We also exploited the deep HST ACS coverage in the COSMOS field to identify sources in the merger phase.
We selected only sources that appear to be in a clear major merger phase, and over-plotted them in Fig.~\ref{fig:simul} (right)
as black stars, in the first three redshift bins 
(above $z\sim1$ it becomes difficult to assess the AGN host morphology).
This selection is not meant to be complete: not all the sources are covered by ACS, and not for all of them it is possible
to recognize the host morphology, due to bright point-like AGN contribution, for example.
However, it is interesting that AGN hosts clearly in merger state tend to cover the highest sSFR and sBHAR range,  
as predicted by simulations. 

 \begin{figure*}
 \begin{center}
 \includegraphics[width=8cm]{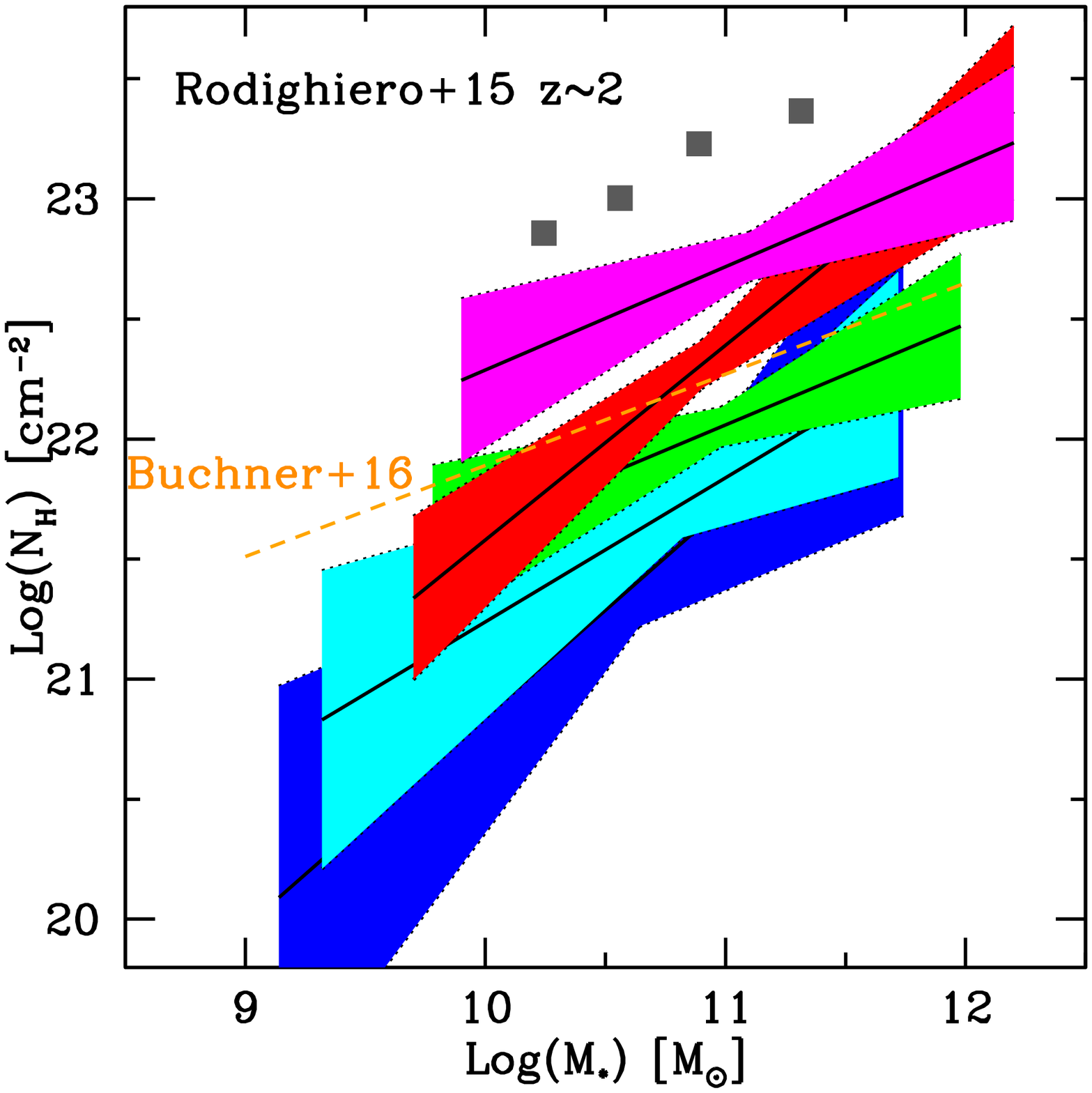}\hspace{1.0cm}\includegraphics[width=8cm]{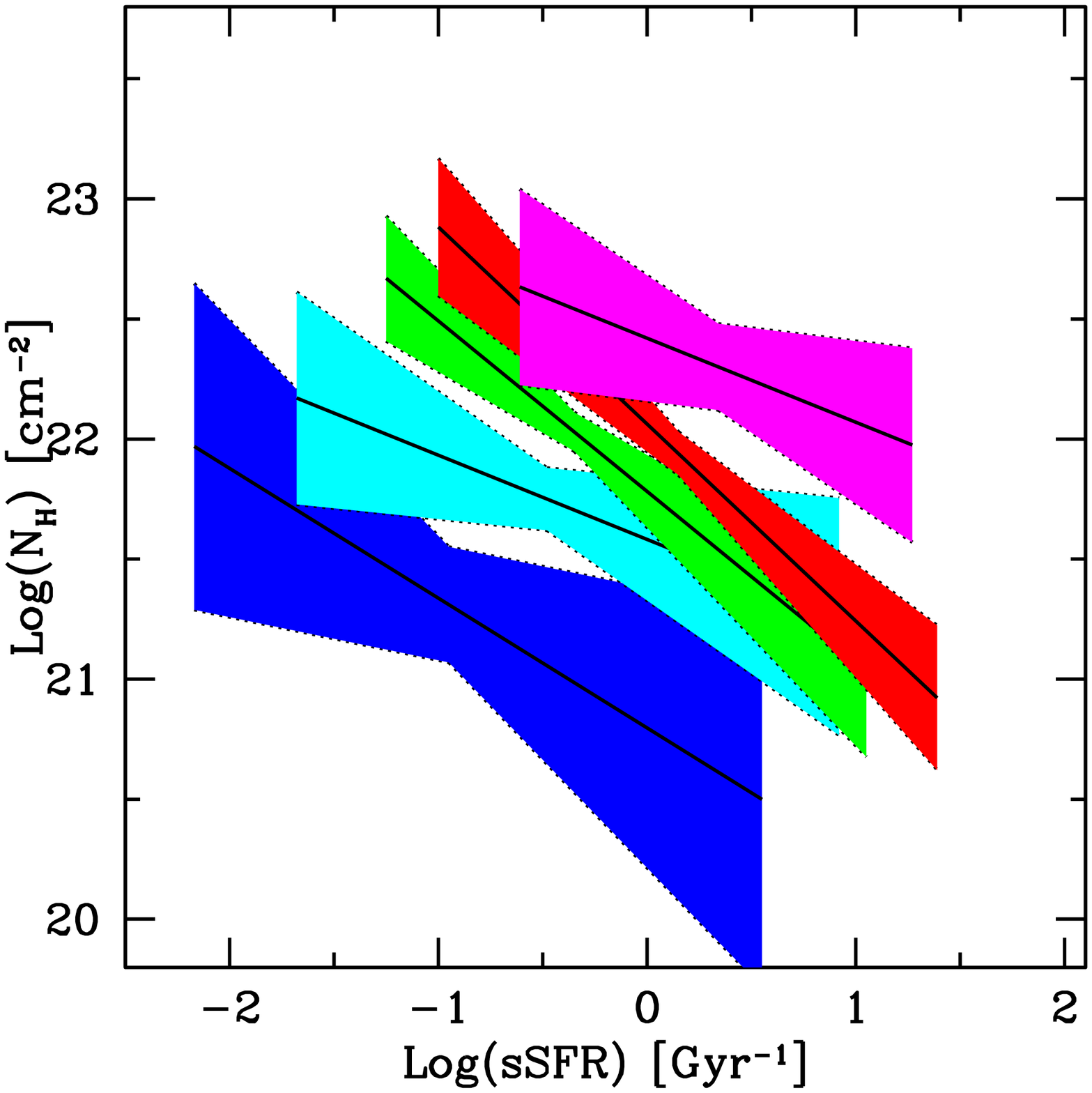}
 \caption{Linear regression of \nh\ vs. \mstar\ (left) and sSFR (right) in five redshift bins.
 The regression is performed using the {\tt linmix} code, that also takes into account the \nh\ upper-limits.
 The color code is the same of Fig. 4.
 The gray squares in the left panel show results from Rodighiero et al. (2015) at $z\sim2$,  
 obtained from the HR of X-ray stacked images of FIR detected galaxies in the COSMOS field. 
 The orange dashed line is the relation found in Buchner et al. (2017) for a sample
 of GRB hosts in a wide range of redshifts (see text).}
 \label{fig:nh_y}
 \end{center}
 \end{figure*}

\subsection{Caveat}

One caveat to be considered here is the fact that the simulations are performed at high-z, starting at $z=3$ 
and ending after $1-3$ Gyr depending on the merger dynamics (see Capelo et al. 2015 for details).
By construction, the simulations have a relatively low gas fraction: 30\% of the disc stellar mass.
This is probably a low value for SF galaxies at these redshifts.
Only one set of simulations has been performed with a higher gas fraction, i.e. 60\% , 
and, as expected, these simulated galaxies move toward higher sSFR and sBHAR as the contours of the 
observed high redshift sample do.

Another caveat is the fact that the simulations are done for low mass galaxies.
The typical \mstar\ for these galaxies is in the range Log(\mstar)$= 9-9.5$ (\msun), 
i.e. in the low mass tail of the mass distribution even for the lowest redshift bin of the observed
sample.
Since the efficiency of SFR and BHAR is most probably mass-dependent, 
the comparison between different mass ranges may not be straightforward. 
Volonteri et al. (2015a) argue, however, that SFR and BHAR are self-similar, 
on the basis of the mass sequence of star forming galaxies and of the possible power-law dependence 
of the specific BHAR (Aird et al. 2012; Bongiorno et al. 2013, but see Kauffmann \& Heckman 2009, 
Lusso et al. 2012 and Schulze et al. 2015). 

Finally, the simulations are not cosmological, in the sense that the gas mass is not replenished 
by cosmic inflows and gas accretion, as it is the case for real galaxies. 
This leads to a possible underestimate of SFR and BHAR towards the end of the simulation, 
when galaxies have converted a large fraction of their gas in stellar and BH mass (see also Vito et al. 2014).

\section{\nh\ and host properties}

 \begin{figure}
 \begin{center}
\includegraphics[width=8.7cm, height=7.8cm]{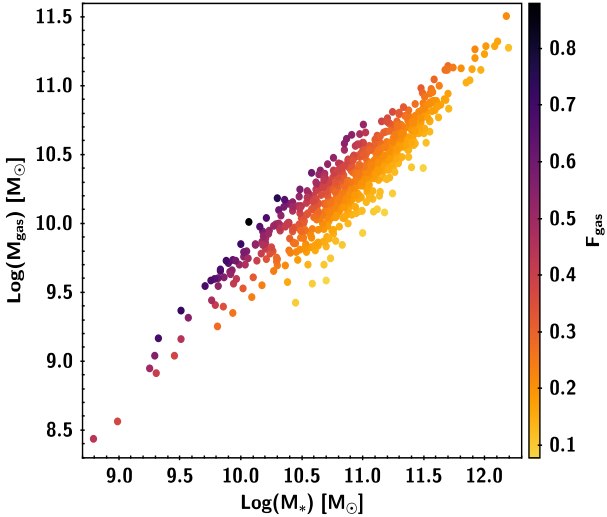}
\caption{$\log$ \mstar\ vs. $\log M_{gas}$ as derived from the eq. 1 of Scoville et al. (2016). 
The sources are color coded on the basis of their gas fraction.}
 \label{fig:gasfr}
 \end{center}
 \end{figure}

Here we discuss the possible correlations between the column density through the AGN line of sight, as measured
by the X-ray \nh, and the host galaxy properties, such as \mstar, SFR and sSFR and MS offset.
The partial correlation analysis described in sec. 3.1, gives a significant positive correlation (at $>4\sigma$ c.l.)
between \nh\ and \mstar, in the entire sample, once the distance effect (both \nh\ and \mstar\ tend to 
increase with redshift in two different ways, due to two different selection effects) is removed.
We also find a significant negative correlation (at $>5\sigma$ c.l.) between \nh\ and sSFR,
while we do not find any significant correlation of \nh\ with SFR and MS offset.

As in the case of \lum\ vs. \lir, the binning direction (or the variable 
chosen as independent) is relevant for the final distribution of 
\nh\ as a function of host properties and vice-versa:
computing average SFR, \mstar, sSFR and MS offset in bin of \nh\ we found
a remarkably flat distribution of all these quantities, 
in agreement with results from Shao et al. (2010), Rovilos et al. (2012), Rosario et al. (2012), where the authors do not
find any evolution of the average host properties in bins of \nh.  

On the other hand, computing average \nh\ values in bins of \mstar\  gives a positive trend in each redshift bin, 
while computing the average \nh\ in sSFR bins gives a negative trend, in agreement with partial correlation analysis. 
The situation in this case is however complicated by the presence of upper-limits in \nh, that makes the problem 
inherently asymmetric. We therefore performed the linear regression of (Y|X)
with a Bayesian approach using the {\tt linmix} code (Kelly et al. 2007) 
that is able to properly take into account the upper-limits on \nh.

The result is shown in Fig.~\ref{fig:nh_y}: the linear regression gives a clear positive correlation 
of \nh\ with the host stellar mass, increasing by one-two dex from low to high masses, at all redshifts
(slopes in the range $\alpha=0.42$ -- $0.88$).
An opposite result is found for the sSFR: the average \nh\ decreases typically by one order of magnitude 
or more, going from low to high sSFR (slopes in the range $\alpha=-0.35$ -- $-0.82$). 
Given that there is no trend of \nh\ with SFR, and that
the sSFR is defined as SFR/\mstar, the two relations are clearly connected.

 \begin{figure*}
 \begin{center}
\includegraphics[width=8cm]{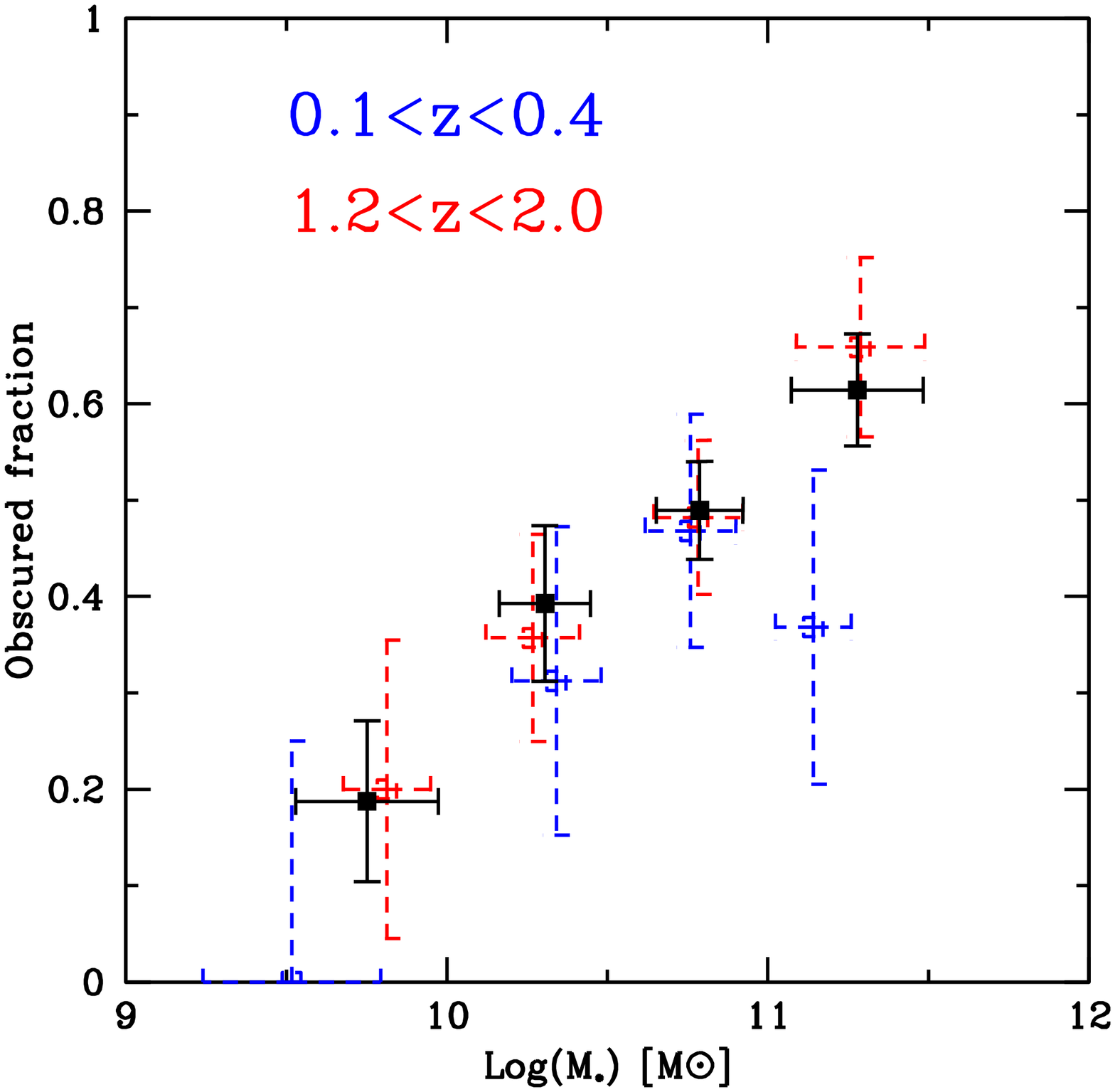}\hspace{1.0cm}\includegraphics[width=8cm]{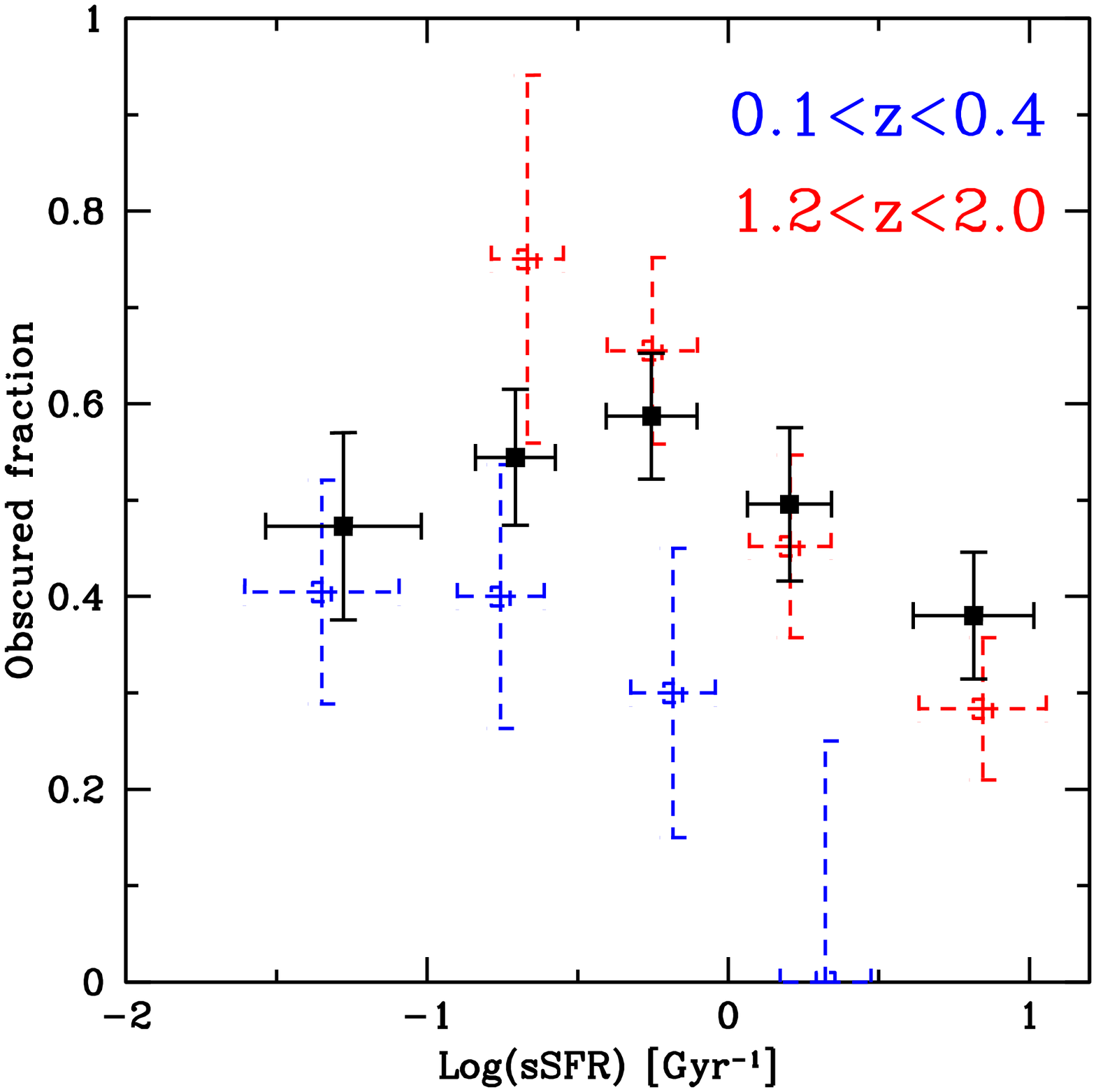}
 \caption{Fraction of obscured sources as a function of \mstar\ (left) and sSFR (right), for the entire 
 sample (black points). The blue (red) dashed points show the results for the first (forth)
 redshift bin, respectively.}
 \label{fig:obsfrac}
 \end{center}
 \end{figure*}

A similar result between \nh\ and \mstar\ was found in Rodighiero et al. (2015) 
for a sample of $z\sim2$ AGN hosts. In their analysis however, the average \nh\ 
is globally $\sim1$ dex higher (see gray squares in Fig.~9, left), 
due to the fact that they derive \nh\ from hardness ratios of the X-ray stacking,
which includes also highly obscured and CT, undetected AGN.

Interestingly, a recent study on the distribution of the obscuration
observed in X-ray spectra of GRB, as a function of the host galaxy mass, found 
a similar trend, in the redshift range $1\simlt z\simlt 5$ (Buchner et al. 2017, orange line in fig. 9 left).
Since for these sources the \nh\ from the GRB spectra is probing only the host obscuration, the authors conclude
that a large fraction of the obscuration observed in AGN, at least in the Compton thin regime,
is not due to the nuclear torus, but to the galaxy-scale gas in the host.

These dependencies imply that at increasing galaxy mass there are more chances to have an additional component 
to the amount of gas and dust along the line of sight through the AGN. 
It is well established that the gas fraction is a strong decreasing function of the galaxy mass 
(e.g. Santini et al. 2014; Peng, Maiolino \& Cochrane 2015). 
However, it is possible to show that the {\it total} amount of gas
is driven mainly by the total galaxy mass, and not by the gas fraction. To this end, 
we computed gas mass for all our galaxies, following the empirical
relation found in Scoville et al. 2016  (their eq. 1), that links \mstar, sSFR offset from the MS, and molecular gas mass.
This is shown in Fig.~\ref{fig:gasfr}, where the sources are color-coded 
on the basis of their gas fraction. 
Even if at increasing \mstar\ the gas fraction is smaller,
the total amount of gas still increases with \mstar.

The well-known mass-metallicity relation (e.g. Tremonti et al. 2004, Mannucci et al. 2010) 
goes in the direction of having more metals
(responsible for X-ray absorption) with increasing \mstar.
In particular, going from Log(\mstar)=9.5 to 11.5,
there is an increase of a factor $\sim2$ in the metallicity, up to $z\sim2$ (Erb et al. 2006). 
This fact is however not enough to explain the increase in average \nh\ observed here:
Measuring the \nh\ with fixed metallicity (as done here) for sources with such a range in
metallicity, translates into a factor $\sim2$ difference in measured \nh, for a given input obscuration.

\subsection{Obscured Fraction}

To compare our results with the literature, we also looked at the fraction of obscured sources
as a function of host properties. In Fig.~\ref{fig:obsfrac} we show the fraction of obscured sources, 
defined as N$_{Obs}$/N$_{Tot}$ where N$_{Obs}$ is the number of sources with a detection of \nh\ and 
\nh$>1\times10^{22}$ cm$^{-2}$. 
As expected from what shown in the previous section, the fraction of obscured sources 
increases with increasing \mstar, and decreases with sSFR (for sSFR$>1$ Gyr$^{-1}$).
The decrease in sSFR is partly washed out by the fact that we are considering 
the full redshift interval ($z=0.1-4$) while Fig.~\ref{fig:nh_y} (right) 
shows that the range covered by the different subsamples shifts toward higher sSFR with redshift.
For this reason we also show in Fig.~\ref{fig:obsfrac} the results for the first and forth bins 
(blue and red dashed points, respectively) as an example.

Merloni et al. (2014) found a flat relation between the fraction of obscured sources
and \mstar\ in a sample of X-ray detected AGN from the XMM-COSMOS catalog.
However, they limited their analysis to a narrow range in \lum\ (in order to cover a wide range redshift),
while the obscured fraction is known to evolve strongly with \lum (e.g. Ueda et al. 2015).

Another group, instead, have found  an increasing fraction of obscured sources
as a function of sSFR and MS offset, in a sample of $70\mu m$ selected galaxies at $0.3<z<1$,
interpreted as an indication of increasing gas fraction or density in the host,
that in turn would sustain the increased sSFR.(e.g. Juneau et al. 2013, J13 hereafter).

We note that the definition of obscured AGN adopted here and in J13 are different,
and in the latter, mostly based on the lack of X-ray detection:
there are 64 sources (out of 99 AGN) classified as obscured AGN on the basis of 
the Mass-Excitation diagram selection (MEX, Juneau et al. 2011), and the X-ray non detection. 
If these objects are indeed highly obscured, Compton-thick AGN, this population is mostly missed in our X-ray based sample.

Another possibility is that a fraction of the MEX-selected AGN are not actively/strongly accreting SMBHs.
Indeed, a sizable fraction ($\sim30\%$) of the AGN selected in J13 through the MEX diagram,
has a host \mstar\ below $Log($\mstar$)=10.5$.
As shown in Sec. 2.2, however, X-ray detected AGN are rare at low \mstar.
Therefore all the sources that are X-ray undetected for reasons different from obscuration 
(variability, intrinsic weakness, contaminant non-AGN etc.) would appear as obscured, low \mstar\ host
(hence high sSFR) AGN, possibly affecting the observed trends.

\section{Discussion}
 
We collected a large sample of X-ray and FIR detected AGN and host systems in the COSMOS
field, spanning $\sim4$ orders of magnitudes in \lum, \nh, \lirsf, \mstar, and covering the redshift range $0.1<z<4$.
We applied X-ray spectral analysis down to very low counts, ($>30$ net counts) and adopted the SED decomposition
results derived in D15, to recover both AGN and SF properties of each source.
With this data-set in hand, we demonstrated that it is possible to reproduce both the flat distribution 
of {\it average} \lirsf\ in bins of \lum\ and the steeper correlation of 
{\it average} \lum\ in bins of \lirsf\ reported in the literature in the latest years 
(e.g. Shao et al. 2010, Rosario et al. 2012, Mullaney et al. 2012, C13, Stanley et al. 2015).

The apparently contradictory results found in the literature, and reproduced in Sec.~3.2,
are due to the different results that are obtained when binning along one axis or the other, 
the equivalent of a forward or inverse linear regression (i.e. \lirsf | \lum vs. \lum | \lirsf), 
as proposed in Hickox et al. (2014) and Volonteri et al. (2015), and found in Dai et al. (2015) on shallow XMM-LSS data.

Both from a physical and a statistical point of view,
it seems more appropriate to consider  the results from \lum | \lirsf, given
the larger measurement uncertainties on \lum,
and the shorter time scale variability of \lum, with respect to \lirsf, 
that adds a further term of intrinsic scatter.
Doing so, we found a linear correlation between \lum\ and \lirsf\ with slope consistent with 1,
at least in the redshift range 0.4-1.2, i.e. below the peak of the SF and BH accretion history.
Beyond that and up to $z=4$, the slope becomes significantly flatter, $\alpha=0.3-0.5$.

The other possibility is to adopt a symmetrical approach, 
even if there is no general agreement on this (see Hogg et al. (2010) on the bisector method). 
In this case the result is a correlation with slope consistent with $\sim1$, at all redshifts.
This would point toward an {\it average} one-to-one 
correlation between SF and BH accretion, in the last 12 Gyr of cosmic history.  

Even more interesting is the full distribution of BH and host properties, such as \lum\ and \lirsf\ or sBHAR and sSFR,
that can be  only qualitatively compared, for the moment, with predictions from galaxy merger simulations, 
resulting in interesting similarities between observations and models.

We stress again that these results apply to the small subsample of 
AGN/host systems detected in both X-ray and FIR, 
that represents only $\sim20$\% of the full X-ray sample and $\sim10$\% of the AGN FIR sample.
Indeed, one of the main reasons why it is so difficult for 
present observations to probe the AGN-SF connection, is the fact that (X-ray and/or FIR) detected systems 
span a limited range in AGN and SF activity, sampling only the high \lum/SFR tail of the possible correlation, 
(e.g. Sijacki et al. 2015).

It is, however, interesting that we are able to reproduce the results obtained via stacking of samples
where the vast majority of the sources are not detected (e.g. 20\% of FIR detected AGN selected in X-ray in Shao et al. 2010).
As suggested in Mullaney et al. (2015), the stacking analysis, being the equivalent of a linear mean, 
may be dominated by the brightest sources. 

A crucial next step in the comparison between theory and observations will be to select the 
observed systems in different evolutionary stage,
to reach a similar level of detail as in the current simulations.
This will be feasible for large samples only at low redshift, while 
detailed and complete morphological studies in COSMOS (and other deep fields) data are very difficult already 
at $z\simgt1$.
From the theoretical point of view, more demanding galaxy merger simulations will be required, 
in order to cover, with the same high resolution, a mass range comparable to the one of observed systems,
and to possibly move toward a high redshift environment.

Finally, a positive correlation between \nh\ and \mstar, and a similar negative correlation 
with sSFR, have been found at all redshift bins. A similar result was found by Rodighiero et al. (2015)
in a large sample of high redshift galaxies, computing HR of stacked X-ray images.
A recent study on GRB hosts has found a similar behavior (Buchner et al. 2017),
implying that an important fraction (up to 40\%) of the Compton thin obscuration found in AGN
can be ascribed to galaxy scale gas (Buchner \& Bauer 2017).

Several studies have found no correlation between column density and host properties 
(Rovilos et al. 2012, Rosario et al. 2012), while others (e.g. J13) have found 
a positive correlation of the fraction of obscured sources with sSFR.
Further investigation in this direction will help to shed light on the role of the host in 
contributing to the obscuration through the AGN line of sight.

\begin{acknowledgements}

The authors thank the anonymous referee for valuable comments.
GL, MB, and MP acknowledge financial support from the CIG grant ``eEASY'' n. 321913.
GL acknowledges financial support from ASI-INAF 2014-045-R.0.
ID acknowledges the European Union’s Seventh Framework programme under grant
agreement 337595 (ERC Starting Grant, ``CoSMas'').
 We acknowledge the contributions of the entire
COSMOS collaboration consisting of more than 100 scientists.
More information on the COSMOS survey is available
at \url{http://cosmos.astro.caltech.edu/}.
Based on observations obtained with XMM-Newton, an ESA science mission with instruments 
and contributions directly funded by ESA Member States and NASA', 
and data obtained from the \chandra\ Data Archive.

\end{acknowledgements}


\begin{appendix}

%
%
%
%
%
%
%

\end{appendix}


\begin{thebibliography}{}


\bibitem[Aird et al.(2012)]{2012ApJ...746...90A} Aird, J., Coil, A.~L., Moustakas, J., et al.\ 2012, \apj, 746, 90 
\bibitem[Akritas \& Bershady(1996)]{1996ApJ...470..706A} Akritas, M.~G., \& Bershady, M.~A.\ 1996, \apj, 470, 706 
\bibitem[Alonso-Herrero et al.(2012)]{2012ApJ...744....2A} Alonso-Herrero, A., Pereira-Santaella, M., Rieke, G.~H., \& Rigopoulou, D.\ 2012, \apj, 744, 2 
\bibitem[Arnaud(1996)]{1996ASPC..101...17A} Arnaud, K.~A.\ 1996, Astronomical Data Analysis Software and Systems V, 101, 17 
\bibitem[Berta et al.(2013)]{2013A&A...551A.100B} Berta, S., Lutz, D., Santini, P., et al.\ 2013, \aap, 551, AA100 
\bibitem[Bennett et al.(2013)]{2013ApJS..208...20B} Bennett, C.~L., Larson, D., Weiland, J.~L., et al.\ 2013, \apjs, 208, 20 
\bibitem[Bongiorno et al.(2012)]{2012MNRAS.427.3103B} Bongiorno, A., Merloni, A., Brusa, M., et al.\ 2012, \mnras, 427, 3103 
\bibitem[Brusa et al.(2009)]{2009A&A...507.1277B} Brusa, M., Fiore, F., Santini, P., et al.\ 2009, \aap, 507, 1277 
\bibitem[Brusa et al.(2010)]{2010ApJ...716..348B} Brusa, M., Civano, F. Comastri, A., et al.\ 2010, \apj, 716, 348 
\bibitem[Bruzual \& Charlo(2003)]{2003MNRAS.344.1000B}Bruzual, G., \& Charlot, S. 2003, \mnras, 344, 1000
\bibitem[Buchner et al.(2017)]{2017MNRAS.464.4545B} Buchner, J., Schulze, S., \& Bauer, F.~E.\ 2017, \mnras, 464, 4545 
\bibitem[Buchner \& Bauer(2017)]{2017MNRAS.465.4348B} Buchner, J., \& Bauer, F.~E.\ 2017, \mnras, 465, 4348 
\bibitem[Bundy et al.(2008)]{2008ApJ...681..931B} Bundy, K., Georgakakis, A., Nandra, K., et al.\ 2008, \apj, 681, 931-943 
\bibitem[Cai et al.(2013)]{2013ApJ...768...21C} Cai, Z.-Y., Lapi, A., Xia, J.-Q., et al.\ 2013, \apj, 768, 21 
\bibitem[Capak et al.(2007)]{2007ApJS..172...99C} Capak, P., Aussel, H., Ajiki, M., et al.\ 2007, \apjs, 172, 99 
\bibitem[Capelo et al.(2015)]{2015MNRAS.447.2123C} Capelo, P.~R., Volonteri, M., Dotti, M., et al.\ 2015, \mnras, 447, 2123 
\bibitem[Chabrier(2003)]{2003ApJ...586L.133C} Chabrier, G.\ 2003, \apjl, 586, L133 
\bibitem[Chen et al.(2013)]{2013ApJ...773....3C} Chen, C.-T.~J., Hickox, R.~C., Alberts, S., et al.\ 2013, C13, \apj, 773, 3 
\bibitem[Chen et al.(2015)]{2015ApJ...802...50C} Chen, C.-T.~J., Hickox, R.~C., Alberts, S., et al.\ 2015, \apj, 802, 50 
\bibitem[Civano et al.(2012)]{2012ApJS..201...30C} Civano, F., Elvis, M., Brusa, M., et al.\ 2012, \apjs, 201, 30 
\bibitem[Civano et al.(2015)]{2015ApJ submitted}  Civano, F., Marchesi, S., Elvis, M., et al. 2015, submitted to \apj
\bibitem[Comastri et al.(2011)]{2011A&A...526L...9C} Comastri, A., Ranalli, P., Iwasawa, K., et al.\ 2011, \aap, 526, L9 
\bibitem[Cresci et al.(2009)]{2009ApJ...697..115C} Cresci, G., Hicks, E.~K.~S., Genzel, R., et al.\ 2009, \apj, 697, 115 
\bibitem[Dai et al.(2015)]{2015arXiv151106761D} Dai, Y.~S., Wilkes, B.~J., Bergeron, J., et al.\ 2015, arXiv:1511.06761 
\bibitem[Delvecchio et al.(2014)]{2014MNRAS.439.2736D} Delvecchio, I., Gruppioni, C., Pozzi, F., et al.\ 2014, \mnras, 439, 2736 
\bibitem[Delvecchio et al.(2015)]{2015arXiv150107602D} Delvecchio, I., Lutz, D., Berta, S., et al.\ 2015, arXiv:1501.07602 
\bibitem[Diamond-Stanic \& Rieke(2012)]{2012ApJ...746..168D} Diamond-Stanic, A.~M., \& Rieke, G.~H.\ 2012, \apj, 746, 168 
\bibitem[Di Matteo et al.(2005)]{2005Natur.433..604D} Di Matteo, T., Springel, V., \& Hernquist, L.\ 2005, \nat, 433, 604 
\bibitem[Dubois et al.(2016)]{2016arXiv160603086D} Dubois, Y., Peirani, S., Pichon, C., et al.\ 2016, arXiv:1606.03086 
\bibitem[Elvis et al.(2009)]{2009ApJS..184..158E} Elvis, M., Civano, F., Vignali, C., et al.\ 2009, \apjs, 184, 158 
\bibitem[Erb et al.(2006)]{2006ApJ...644..813E} Erb, D.~K., Shapley, A.~E., Pettini, M., et al.\ 2006, \apj, 644, 813 
\bibitem[Fabian \& Iwasawa(1999)]{1999MNRAS.303L..34F} Fabian, A.~C., \& Iwasawa, K.\ 1999, \mnras, 303, L34 
\bibitem[Feigelson \& Berg(1983)]{1983ApJ...269..400F} Feigelson, E.~D., \& Berg, C.~J.\ 1983, \apj, 269, 400 
\bibitem[Feltre et al.(2012)]{2012MNRAS.426..120F} Feltre, A., Hatziminaoglou, E., Fritz, J., \& Franceschini, A.\ 2012, \mnras, 426, 120 
\bibitem[Fritz et al.(2006)]{2006MNRAS.366..767F} Fritz, J., Franceschini, A., \& Hatziminaoglou, E.\ 2006, \mnras, 366, 767 
\bibitem[Fruscione et al.(2006)]{2006SPIE.6270E..60F} Fruscione, A., et al.\ 2006, \procspie, 6270
\bibitem[Genzel et al.(2011)]{2011ApJ...733..101G} Genzel, R., Newman, S., Jones, T., et al.\ 2011, \apj, 733, 101 
\bibitem[Granato et al.(2004)]{2004ApJ...600..580G} Granato, G.~L., De Zotti, G., Silva, L., Bressan, A., \& Danese, L.\ 2004, \apj, 600, 580 
\bibitem[Griffin et al.(2010)]{2010A&A...518L...3G} Griffin, M.~J., Abergel, A., Abreu, A., et al.\ 2010, \aap, 518, L3 
\bibitem[Gruppioni et al.(2016)]{2016MNRAS.458.4297G} Gruppioni, C., Berta, S., Spinoglio, L., et al.\ 2016, \mnras, 458, 4297 
\bibitem[Hasinger et al.(2007)]{2007ApJS..172...29H} Hasinger, G., Cappelluti, N., Brunner, H., et al.\ 2007, \apjs, 172, 29 
\bibitem[Hickox et al.(2014)]{2014ApJ...782....9H} Hickox, R.~C., Mullaney, J.~R., Alexander, D.~M., et al.\ 2014, \apj, 782, 9 
\bibitem[Hogg et al.(2010)]{2010arXiv1008.4686H} Hogg, D.~W., Bovy, J., \& Lang, D.\ 2010, arXiv:1008.4686 
\bibitem[Imanishi et al.(2010)]{2010ApJ...709..801I} Imanishi, M., Maiolino, R., \& Nakagawa, T.\ 2010, \apj, 709, 801 
\bibitem[Isobe et al.(1990)]{1990ApJ...364..104I} Isobe, T., Feigelson, E.~D., Akritas, M.~G., \& Babu, G.~J.\ 1990, \apj, 364, 104 
\bibitem[Juneau et al.(2011)]{2011ApJ...736..104J} Juneau, S., Dickinson, M., Alexander, D.~M., \& Salim, S.\ 2011, \apj, 736, 104 
\bibitem[Juneau et al.(2013)]{2013ApJ...764..176J} Juneau, S., Dickinson, M., Bournaud, F., et al.\ 2013, J13, \apj, 764, 176 
\bibitem[Kauffmann et al.(2003)]{2003MNRAS.346.1055K} Kauffmann, G., Heckman, T.~M., Tremonti, C., et al.\ 2003, \mnras, 346, 1055 
\bibitem[Kauffmann \& Heckman(2009)]{2009MNRAS.397..135K} Kauffmann, G., \& Heckman, T.~M.\ 2009, \mnras, 397, 135 
\bibitem[Kelly(2007)]{2007ApJ...665.1489K} Kelly, B.~C.\ 2007, \apj, 665, 1489 
\bibitem[Kennicutt(1998)]{1998ApJ...498..541K} Kennicutt, R.~C., Jr.\ 1998, \apj, 498, 541 
\bibitem[Koekemoer et al.(2007)]{2007ApJS..172..196K} Koekemoer, A.~M., Aussel, H., Calzetti, D., et al.\ 2007, \apjs, 172, 196 
\bibitem[Kormendy \& Richstone(1995)]{1995ARA&A..33..581K} Kormendy, J., \& Richstone, D.\ 1995, \araa, 33, 581 
\bibitem[Kormendy \& Ho(2013)]{2013ARA&A..51..511K} Kormendy, J., \& Ho, L.~C.\ 2013, \araa, 51, 511 
\bibitem[Lagos et al.(2011)]{2011MNRAS.418.1649L} Lagos, C.~D.~P., Baugh, C.~M., Lacey, C.~G., et al.\ 2011, \mnras, 418, 1649 
\bibitem[Lanzuisi et al.(2013)]{2013MNRAS.431..978L} Lanzuisi, G., Civano, F., Elvis, M., et al.\ 2013, \mnras, 431, 978 
\bibitem[Lanzuisi et al.(2015)]{2015A&A...573A.137L} Lanzuisi, G., Ranalli, P., Georgantopoulos, I., et al.\ 2015a, \aap, 573, AA137 (L15)
\bibitem[Lanzuisi et al.(2015)]{2015A&A...578A.120L} Lanzuisi, G., Perna, M., Delvecchio, I., et al.\ 2015b, \aap, 578, A120 
\bibitem[Lusso et al.(2012)]{2012MNRAS.425..623L} Lusso, E., Comastri, A., Simmons, B.~D., et al.\ 2012, \mnras, 425, 623 (L12)
\bibitem[Lutz et al.(1998)]{1998ApJ...505L.103L} Lutz, D., Spoon, H.~W.~W., Rigopoulou, D., Moorwood, A.~F.~M., \& Genzel, R.\ 1998, \apjl, 505, L103 
\bibitem[Lutz et al.(2011)]{2011A&A...532A..90L} Lutz, D., Poglitsch, A., Altieri, B., et al.\ 2011, \aap, 532, A90 
\bibitem[Macklin(1982)]{1982MNRAS.199.1119M} Macklin, J.~T.\ 1982, \mnras, 199, 1119 
\bibitem[Madau \& Dickinson(2014)]{2014ARA&A..52..415M} Madau, P., \& Dickinson, M.\ 2014, \araa, 52, 415 
\bibitem[Magorrian et al.(1998)]{1998AJ....115.2285M} Magorrian, J., Tremaine, S., Richstone, D., et al.\ 1998, \aj, 115, 2285 
\bibitem[Mainieri et al.(2011)]{2011A&A...535A..80M} Mainieri, V., Bongiorno, A., Merloni, A., et al.\ 2011, \aap, 535, A80 
\bibitem[Maiolino et al.(2007)]{2007A&A...468..979M} Maiolino, R., Shemmer, O., Imanishi, M., et al.\ 2007, \aap, 468, 979 
\bibitem[Mannucci et al.(2010)]{2010MNRAS.408.2115M} Mannucci, F., Cresci, G., Maiolino, R., Marconi, A., \& Gnerucci, A.\ 2010, \mnras, 408, 2115 
\bibitem[Marchesi et al.(2016)]{2016ApJ...817...34M} Marchesi, S., Civano, F., Elvis, M., et al.\ 2016, \apj, 817, 34 
\bibitem[Marchesi et al.(2016)]{2016ApJ...830..100M} Marchesi, S., Lanzuisi, G., Civano, F., et al.\ 2016, \apj, 830, 100 
\bibitem[Marconi \& Hunt(2003)]{2003ApJ...589L..21M} Marconi, A., \& Hunt, L.~K.\ 2003, \apjl, 589, L21 
\bibitem[Marconi et al.(2004)]{2004MNRAS.351..169M} Marconi, A., Risaliti, G., Gilli, R., et al.\ 2004, \mnras, 351, 169 
\bibitem[Menci et al.(2008)]{2008ApJ...686..219M} Menci, N., Fiore, F., Puccetti, S., \& Cavaliere, A.\ 2008, \apj, 686, 219 
\bibitem[Merloni et al.(2014)]{2014MNRAS.437.3550M} Merloni, A., Bongiorno, A., Brusa, M., et al.\ 2014, \mnras, 437, 3550 
\bibitem[Mullaney et al.(2011)]{2011MNRAS.414.1082M} Mullaney, J.~R., Alexander, D.~M., Goulding, A.~D., \& Hickox, R.~C.\ 2011, \mnras, 414, 1082 
\bibitem[Mullaney et al.(2012)]{2012MNRAS.419...95M} Mullaney, J.~R., Pannella, M., Daddi, E., et al.\ 2012, \mnras, 419, 95 
\bibitem[Mullaney et al.(2015)]{2015MNRAS.453L..83M} Mullaney, J.~R., Alexander, D.~M., Aird, J., et al.\ 2015, \mnras, 453, L83 
\bibitem[Neistein \& Netzer(2014)]{2014MNRAS.437.3373N} Neistein, E., \& Netzer, H.\ 2014, \mnras, 437, 3373 
\bibitem[Novak et al.(2006)]{2006ApJ...637...96N} Novak, G.~S., Faber, S.~M., \& Dekel, A.\ 2006, \apj, 637, 96 
\bibitem[Oliver et al.(2012)]{2012MNRAS.424.1614O} Oliver, S.~J., Bock, J., Altieri, B., et al.\ 2012, \mnras, 424, 1614 
\bibitem[Peng et al.(2015)]{2015Natur.521..192P} Peng, Y., Maiolino, R., \& Cochrane, R.\ 2015, \nat, 521, 192 
\bibitem[Pilbratt et al.(2010)]{2010A&A...518L...1P} Pilbratt, G.~L., Riedinger, J.~R., Passvogel, T., et al.\ 2010, \aap, 518, L1 
\bibitem[Poglitsch et al.(2010)]{2010A&A...518L...2P} Poglitsch, A., Waelkens, C., Geis, N., et al.\ 2010, \aap, 518, L2 
\bibitem[Pontzen et al.(2016)]{2016arXiv160702507P} Pontzen, A., Tremmel, M., Roth, N., et al.\ 2016, arXiv:1607.02507 
\bibitem[Pozzi et al.(2012)]{2012MNRAS.423.1909P} Pozzi, F., Vignali, C., Gruppioni, C., et al.\ 2012, \mnras, 423, 1909 
\bibitem[Rodighiero et al.(2011)]{2011ApJ...739L..40R} Rodighiero, G., Daddi, E., Baronchelli, I., et al.\ 2011, \apjl, 739, L40 
\bibitem[Rodighiero et al.(2015)]{2015ApJ...800L..10R} Rodighiero, G., Brusa, M., Daddi, E., et al.\ 2015, \apjl, 800, L10 
\bibitem[Rosario et al.(2012)]{2012A&A...545A..45R} Rosario, D.~J., Santini, P., Lutz, D., et al.\ 2012, \aap, 545, A45 
\bibitem[Rovilos et al.(2012)]{2012A&A...546A..58R} Rovilos, E., Comastri, A., Gilli, R., et al.\ 2012, \aap, 546, A58 
\bibitem[Saintonge et al.(2016)]{2016arXiv160705289S} Saintonge, A., Catinella, B., Cortese, L., et al.\ 2016, arXiv:1607.05289 
\bibitem[Salvato et al.(2009)]{2009ApJ...690.1250S} Salvato, M., Hasinger, G., Ilbert, O., et al.\ 2009, \apj, 690, 1250 
\bibitem[Santini et al.(2012)]{2012A&A...540A.109S} Santini, P., Rosario, D.~J., Shao, L., et al.\ 2012, \aap, 540, A109 
\bibitem[Santini et al.(2014)]{2014A&A...562A..30S} Santini, P., Maiolino, R., Magnelli, B., et al.\ 2014, \aap, 562, A30 
\bibitem[Schulze et al.(2015)]{2015MNRAS.447.2085S} Schulze, A., Bongiorno, A., Gavignaud, I., et al.\ 2015, \mnras, 447, 2085 
\bibitem[Scoville et al.(2007)]{2007ApJS..172....1S} Scoville, N., Aussel, H., Brusa, M., et al.\ 2007, \apjs, 172, 1 
\bibitem[Scoville et al.(2016)]{2016ApJ...820...83S} Scoville, N., Sheth, K., Aussel, H., et al.\ 2016, \apj, 820, 83 
\bibitem[Scudder et al.(2016)]{2016MNRAS.460.1119S} Scudder, J.~M., Oliver, S., Hurley, P.~D., et al.\ 2016, \mnras, 460, 1119 
\bibitem[Shao et al.(2010)]{2010A&A...518L..26S} Shao, L., Lutz, D., Nordon, R., et al.\ 2010, \aap, 518, L26 
\bibitem[Sijacki et al.(2015)]{2015MNRAS.452..575S} Sijacki, D., Vogelsberger, M., Genel, S., et al.\ 2015, \mnras, 452, 575 
\bibitem[Silverman et al.(2009)]{2009ApJ...696..396S} Silverman, J.~D., Lamareille, F., Maier, C., et al.\ 2009, \apj, 696, 396 
\bibitem[Suh et al. (2017 sub.)]{2017ApJ submitted} Suh, H. Civano, F., Hasinger, G., et al.\ 2017, submitted
\bibitem[Symeonidis et al.(2016)]{2016MNRAS.459..257S} Symeonidis, M., Giblin, B.~M., Page, M.~J., et al.\ 2016, \mnras, 459, 257 
\bibitem[Tremaine et al.(2002)]{2002ApJ...574..740T} Tremaine, S., Gebhardt, K., Bender, R., et al.\ 2002, \apj, 574, 740 
\bibitem[Tremonti et al.(2004)]{2004ApJ...613..898T} Tremonti, C.~A., Heckman, T.~M., Kauffmann, G., et al.\ 2004, \apj, 613, 898 
\bibitem[Ueda et al.(2014)]{2014ApJ...786..104U} Ueda, Y., Akiyama, M., Hasinger, G., Miyaji, T., \& Watson, M.~G.\ 2014, \apj, 786, 104 
\bibitem[Vito et al.(2014)]{2014MNRAS.441.1059V} Vito, F., Maiolino, R., Santini, P., et al.\ 2014, \mnras, 441, 1059 
\bibitem[Volonteri et al.(2015)]{2015MNRAS.449.1470V} Volonteri, M., Capelo, P.~R., Netzer, H., et al.\ 2015a, \mnras, 449, 1470 
\bibitem[Volonteri et al.(2015)]{2015MNRAS.452L...6V} Volonteri, M., Capelo, P.~R., Netzer, H., et al.\ 2015b, \mnras, 452, L6 
\bibitem[Whitaker et al.(2012)]{2012ApJ...754L..29W} Whitaker, K.~E., van Dokkum, P.~G., Brammer, G., \& Franx, M.\ 2012, \apjl, 754, LL29 
\bibitem[Xue et al.(2010)]{2010ApJ...720..368X} Xue, Y.~Q., Brandt, W.~N., Luo, B., et al.\ 2010, \apj, 720, 368 


\end{thebibliography}
\end{document}